\documentclass[11pt,preprint]{aastex}

\usepackage{natbib,pstcol}

\slugcomment{Submitted to the Astrophysical Journal}
\shortauthors{Sipior, Eracleous, \& Sigurdsson}
\shorttitle{Evolution of HMXBs from a Starburst}

\newcommand\ergs{\mbox{erg~s$^{-1}$}}
\newcommand\kmsec{\mbox{km~s$^{-1}$}}
\newcommand\Msun{\mbox{{\rm M}$_{\odot}$}}
\newcommand\Msunyr{\mbox{\Msun~yr$^{-1}$}}

\newcommand{\nsns}{NS-NS~}

\shorttitle{X-Rays from Young Stellar Populations}
\shortauthors{Sipior, Eracleous, \& Sigurdsson}

\begin{document}

\title{Simulating the Early Evolution of the Hard X-Ray Properties of
a Young Stellar Population\label{model}}

\author{Michael S. Sipior\altaffilmark{1}, Michael Eracleous, \&
Steinn Sigurdsson}

\affil{Department of Astronomy and Astrophysics, The Pennsylvania State
University, 525 Davey Lab, University Park, PA~16802}

\email{sipior@science.uva.nl, mce@astro.psu.edu, steinn@astro.psu.edu}

\altaffiltext{1}{Current address: Astronomical Institute ``Anton
Pannekoek'' and Section Computational Science, University of
Amsterdam, Kruislaan 403, 1098 SJ Amsterdam, Netherlands.}

\begin{abstract}
We present an X-ray binary population synthesis model, and use it to
simulate the evolution of X-ray binaries formed in a burst of star
formation of duration 20~Myr and star-formation rate
10~\Msun~yr$^{-1}$.  Our goal is to explain the hard (2--10~keV) X-ray
properties of populations of extragalactic X-ray binaries recently
observed by the {\it Chandra} X-Ray Observatory, especially those
associated with recent or ongoing episodes of vigorous star formation.
Our simulated X-ray binary population reaches a maximum 2--10~keV
luminosity of $\sim 4\times10^{40}$~\ergs\ after approximately 20~Myr.
The X-ray luminous phase is sustained for a period of several hundreds
of Myr by succeeding populations of systems with lighter secondary
stars, i.~e., it persists long after the star-formation episode has
ended.  These results are insensitive to the poorly-constrained values
of the initial mass function and the average mass ratio between
accreting and donor stars. The computed peak X-ray luminosity is
consistent with observationally-derived correlations between the
star-formation rate and the observed hard X-ray luminosity. Model
cumulative luminosity functions at the earliest times have power-law
indices in agreement with those derived from observations of actively
star-forming galaxies. The model cumulative luminosity functions
become increasingly steeper with time as the most luminous systems die
off, which offers an explanation for the difference in the slopes of
observed cumulative luminosity functions of young and old stellar
populations.
\end{abstract}


\section{Introduction\label{model:intro}}

\subsection{X-Ray Binaries Revealed By \emph{Chandra}}

There is now a considerable corpus of evidence that, for ``normal''
galaxies (i.~e. with no active nucleus), the principal component of
the X-ray luminosity above 2~keV arises from the associated population
of X-ray binaries (XRBs). This is especially true in the presence of
vigorous starburst activity, and has been further established with the
advent of the \emph{Chandra} X-ray Observatory, where high-resolution
imaging allows an accurate source census and luminosity function to be
constructed for a diverse sample of host galaxies. Specific examples
of large, luminous XRB populations that have been revealed with
\emph{Chandra} include the vigorous starbursts in NGC~4038/4039
\citep[the Antennae,][]{fabbiano:2001}, M82
\citep{zezas:2002,griffiths:2000}, and the ULIRG NGC~3256
\citep{lira:2002}.  Perhaps more interesting, however, has been the
discovery of sizable XRB populations in non-starbursting galaxies,
with inactive or mildly-active galactic nuclei, comparable to those
found in some starbursts. The implication, given the generally slow
star-formation rate (SFR), is that a significant fraction of X-ray
binaries can remain luminous for many gigayears affecting the X-ray
emission of a galaxy long after their formation.  By way of example,
NGC~1291 \citep{irwin:2002}, IC~5332, M83 \citep{kilgard:2002}, and
NGC~4736 \citep{eracleous:2002} all sport several dozen point sources
with 2--10 keV X-ray luminosities well in excess of
$10^{37}\;\mbox{erg s}^{-1}$. The first two of these are
undistinguished spirals; NGC~1291 has a current SFR several times
lower than that seen in the Milky Way, at around 0.1 \Msunyr
\citep{caldwell:1991}, and yet the two galaxies have roughly
comparable populations of luminous XRBs \citep{grimm:2003}, implying
that their historical SFRs were similar.  M83 is known
to exhibit highly localized starburst activity in the nucleus and bar
regions \citep{telesco:1993}, though the observed XRB population is
not limited to these areas.  The nucleus of NGC~4736 is known to
contain a LINER (Low-Ionization Nuclear Emission Region; see
\citealt{heckman:1980}). Although some LINERs are powered by accretion
onto a nuclear, supermassive black hole occupying the low end of the
spectrum of nuclear activity, a significant fraction of them (perhaps
the majority) are weak, compact starbursts. Such LINERs thus occupy
the low end of the spectrum of starburst activity.

More than any other quantity observable in the X-ray band, the
luminosity distribution of an XRB population gives great insight into
the recent star formation history of the host galaxy. Measurements of
soft X-ray superwinds provide a great deal of information on
\emph{recent} episodes of star formation; however, XRBs are more
durable records of the recent past, and can be studied long after the
star formation driving the superwind has faded. This may be
particularly useful in galaxies with moderate SFRs (such as some
LINERs), where a measurable superwind may simply never form.

With the above observational considerations in mind, we embarked on a
theoretical study of the integrated hard X-ray properties of a young
stellar population. Our primary goal was to simulate the formation and
evolution of interacting binaries and to compute the evolution of the
integrated X-ray luminosity as well as the statistical properties of
the X-ray source population. The results of our study are the subject
of this paper.

\subsection{The Goals of Our Simulations}

The immediate goal of the simulations undertaken here is to predict
the evolution of the integrated hard X-ray luminosity of a young
stellar population, formed in a recent episode of star formation, as
well as the evolution of the luminosity function of its discrete X-ray
sources. Therefore, we concentrate on the relatively simple problem of
simulating a brief episode of steady star formation. Later work may
focus on particular case studies in which individual galaxies are
modeled using observational information on their SFRs and
star-formation histories.

The evolution of the discrete-source luminosity function is of
particular interest since recent observations show that it is closely
related to recent or current star-formation activity.  The analyses of
\citet{eracleous:2002} and \citet{kilgard:2002} have shown that the
luminosity function in starburst galaxies tends to be substantially
flatter, with high-luminosity sources in greater abundance.  Moreover,
the slope of the luminosity function shows a strong correlation to the
observed 60~$\mu$m and H$\alpha$ luminosities, direct measures of star
formation \citep{kennicutt:1983,lonsdale:1987,devereux:1990}.  This
trend can be qualitatively understood in terms of the evolutionary
time scales of various components of the XRB population. For low-mass
X-ray binaries (LMXBs), mass transfer is driven by either the nuclear
evolution of the donor or the loss of angular momentum through
gravitational radiation and magnetic braking processes.  All of these
mechanisms operate over relatively long time scales. In contrast,
high-mass X-ray binaries (HMXBs) begin mass transfer on the shorter
nuclear time scale of the massive donor star. HMXBs powered by
Roche-lobe overflow tend to be brighter, on average, than LMXBs, as
the accretor is more likely to be a black hole when the donor is
massive. The result is a flat luminosity function (more numerous
bright X-ray sources) for young stellar populations, which slowly
becomes steeper as the short-lived HMXBs give way to long-lived
LMXBs. The quantitative investigation of the above processes is an
important goal of the work presented here.

The problem of the long-term evolution of XRBs and its observational
signatures has recently been approached in a semi-numerical fashion by
\citet{ghosh:2001,ptak:2001b,white:1998}, in addition to a completely
analytic formalism presented in \citet{wu:2001}.  Our goal is to
approach the study of the evolution of galactic X-ray properties from
a numerical standpoint, which requires fewer simplifying assumptions
than a semi-numerical or purely analytic formulation.

\section{Population Modeling\label{model:modeling}}

\subsection{History and Authorship of the Population Synthesis Code}

We make use of a binary evolution code detailed in part by
\citet{pols:1994}, and modified for use in neutron star-neutron star
(\nsns) systems by \citet{bloom:1999}. Our extension of the code
allows for evolution to the black hole state, with assumptions about
the mass function of such objects at the time of collapse; in
addition, the technique for computing mass transfer rates was refined
considerably, by coupling it more directly to the underlying physics,
as discussed below.

For purposes of the numerical model, an initial binary system is
considered to be completely described by four parameters: the mass of
the system's primary (more massive) star, $M_1$, which is chosen from
the specified initial mass function; the initial primary to secondary
mass ratio, $q$, defined to lie between zero and unity; the initial
orbital eccentricity, $e$; and the initial orbital semi-major axis,
$a$. The code then evolves the binary in time, taking into account the
orbital changes caused by mass transfer, wind loss, etc. These four
quantities are tracked, as is the evolutionary state of each star, and
any mass exchanges that take place. The simulation of the evolution of
the binary ends when both stars have reached their respective
evolutionary end points, which are a strict function of the initial
core mass.

\subsection{Population Code Theory of Operation, Choice and Extent of 
Parameter Space}

The evolution of a binary pair starts with the choice of a primary
mass from an assumed initial mass function (IMF). For this work, we
consider two power law IMFs (where $dN = m^{-\alpha}dm$); the first index
is $\alpha = -2.35$ (the Salpeter IMF; \citealt{salpeter:1955}), the
second is $\alpha = -2.7$, approximating the high end of a Miller-Scalo
IMF \citep{miller:1979}.  In both cases, we established a lower cutoff
of 4~\Msun\ for the primary star's mass, confining the code
to an interesting range of initial masses; i.~e., where at least one
supernova is possible in principle. This is because our interest is in
systems with a neutron star or black hole, as these are the potentially
luminous X-ray sources. Our stellar models are taken primarily from
\citet{maeder:1989}. The helium star models used are a mix of models from
\citet{habets:1986} and \citet{pacz:1971}, and the reader is referred to
these for a detailed discussion and evolutionary tracks.

The distribution of $q$ is a topic of some controversy, given the
observational biases involved in studying systems with diverse mass
ratios \citep{hogeveen:1992}. A ``flat'' distribution, where all
values of $q$ are equally likely, is often chosen given the difficulty
in reconstructing the underlying function. An extensive inventory of
observational data was compiled by \citet{kuiper:1935b} in an attempt
to address the mass ratio distribution question. These data, coupled
with the more recent data of \citet{batten:1989}, and the analysis
found in \citet{hogeveen:1992}, point to two principal results. First,
in the case of single-lined spectroscopic binaries, the distribution
of $q$ is a two-part function, where:
\begin{equation}
\label{eqn:q}
\psi(q) \propto \left\{ \begin{array}{c@{\quad \mbox{for} \quad}l} 
q^{-2} & q > 0.3 \\ 1 & q < 0.3\end{array}\right.
\end{equation}
For double-lined spectroscopic systems, the observed $q$ distribution
was found to be driven almost completely by selection effects, albeit
consistent with the $q$-distribution of single-lined binaries
above. See \citet{elson:1998} for a further discussion of this problem
in the context of massive binaries in a young LMC cluster, where a $q$
distribution biased towards companions of equal mass is found, but the
detection limit prevents an accurate census of low-$q$ systems.  We
consider both the flat and the low-skewed $q$-distributions in our
simulations below, accepting that reality likely lies somewhere
between these two points.

The initial binary separation is chosen after \citet{abt:1983}, with a
distribution that is flat in the logarithm of the semi-major axis ,
and in the range $10 \,{\rm R}_{\odot} < a < 10^6 \,{\rm
R}_{\odot}$. This distribution fits well with existing spectroscopic
surveys of nearby stars.  \citet{duquennoy:1991}, describe another
separation distribution, based upon a CORAVEL spectroscopic survey of
181 Gliese catalogue stars. The function they derive is Gaussian, with
a mean of $8\times10^3\;{\rm R}_{\odot}$ and a dispersion of
$8\times10^2\;{\rm R}_{\odot}$. We use Abt's prescription here, but
wished to make clear that this is not a settled issue. Related
population synthesis studies currently use the former distribution
almost exclusively, often with little comment. If the
\citet{duquennoy:1991} result holds when expanded to a larger survey
size (preferably including a few non-local systems) then this issue
will have to be revisited. Given the dramatically wider initial
separations implied by the \citet{duquennoy:1991} distribution, one
can at least make the prediction that far fewer X-ray binaries would
result, since the common-envelope phase would be less likely to
occur. This in turn would imply wider systems with larger Roche
surfaces, making Roche-lobe overflow less likely.

The eccentricity is chosen from the standard thermal distribution, $\xi(e)
= 2e$. This choice for the distribution of initial eccentricities is
ubiquitous in binary population synthesis. The mathematics justifying
this relation can be found in \citet{heggie:1975}, and interested readers
are referred there for all of the details. 

After the initial parameters have been selected, each binary system is
evolved along the stellar tracks referenced above until both
components have reached their final degenerate form, accounting for
mass-transfer-induced stellar regeneration and stellar winds. Stellar
winds from helium stars are accounted for using the relation developed
in \citet{langer:1989}, where the mass loss rate is $\dot{M} = 5
\times 10^{-8}\; ({M/\Msun})^{2.5} \;\mbox{\rm M}_{\odot}\;\mbox{\rm yr}^{-1}$.
This wind lasts for the duration of the star's helium main sequence
lifetime.

When the more massive primary leaves the main sequence and ascends the
giant branch, the rapidly-swelling star may engulf its companion with
its outer envelope. This common-envelope phase will rapidly shrink the
orbital radius of the binary on a time scale of only a few orbital
periods. Those systems that avoid a merger event at the end of the
common-envelope phase will be more likely to engage in mass transfer,
as the size of the companion's Roche lobe shrinks along with the
orbital separation. A common-envelope phase can also result as the
secondary leaves the main sequence, though systems are unlikely to
survive two such events without merging. For our purposes here, during
the common-envelope phase, the orbit is circularized, and the orbital
energy is reduced by the binding energy of the envelope divided by the
common-envelope efficiency parameter, which we take to be 0.5.  In
other words, the orbital energy is reduced by twice the envelope
binding energy.  The common-envelope efficiency is just the fraction
of the binary orbital energy which is required to eject the common
envelope. The value of this quantity is contentious at present, as the
microphysics necessary to model such an environment is poorly
understood. Attempts at estimating this parameter can be found in
\citet{livio:1988}, \citet{rasio:1996} and \citet{dewi:2000}, among
others.

Neutron stars are formed from progenitors with zero-age main sequence
(ZAMS) masses of between 8 and 20~\Msun, inclusive, and are
always given a mass of 1.4~\Msun. More massive stars end up as
black holes. This boundary is unlikely to be a sharp one, as it is
strongly coupled to the spin state of the pre-collapse object
\citep{fryer:1999}. Even assuming this was known to perfect accuracy,
the effects of magnetic fields and rotational support on the compact
object's end state are not well understood. This point also bears upon
the magnitude and direction distribution of natal kicks received by
the neutron star at birth, from an asymmetric emission of neutrinos or
core material. The role of and justification for asymmetric natal
kicks is discussed below, including the appearance of asymmetric kicks
during black hole formation.

The black hole mass function (i.~e., the post-collapse mass of a black
hole, given its mass just prior to the explosion) is highly
speculative at this point, and is almost certainly not merely a
function of initial mass, but also of angular momentum, to the extent
that this determines the fraction of material falling back onto the
collapsing star. In order to experience a kick, the black hole's
formation must be delayed somewhat, either due to rotational support,
or because event horizon formation occurs only after delayed fallback
of mass initially ejected from the core. \citet{fryer:1999} has
performed core-collapse simulations in order to explore the critical
mass for black hole formation, and the final masses of the resulting
black holes. As a best working scenario, we have constructed a mass
relation from a quadratic fit to the limited data set found in
\citet{fryer:1999}. Our fit shows that the mass of the black hole at
formation ($M_{BH}$) is related to the ZAMS mass of the progenitor
($M_0$) by $M_{BH} = \left(M_0/25\;{\rm M}_{\odot}\right)^2 \times
5.2\;{\rm M}_{\odot}$.  This relation is accurate to about 10\% of the
black hole initial mass at each of the values resulting from a
hydrodynamic simulation.  It should be noted that this relation is
almost certainly dependent upon metallicity (see, for example,
\citealt{fryer:2002}). The assumed relation is appropriate for systems
with approximately solar metallicity, but would need to be adjusted to
represent metal-poor progenitors.

The importance of the black hole IMF in XRB formation extends beyond
the obvious effect on a system's orbital elements. The maximum
luminosity from future mass transfer onto the black hole is a function
of the hole mass.  The Eddington limit,
$L_{Edd} = 1.3\times10^{38}\; (M/{\rm M_{\odot})~erg~s^{-1}}$,
is frequently invoked for this luminosity cutoff; however, this limit
applies strictly only in the case of spherically symmetric accretion.
The black hole IMF is important because it is one of two important
factors that set the maximum luminosity of the most luminous XRBs (the
other is the range of mass transfer rates from the companion star).
Thus, the high-luminosity end of the resulting luminosity function
depends quite sensitively on the black hole IMF.

Of course, the Eddington assumption can be relaxed, allowing us to test the
models that have been put forward to explain the significant number of
extremely-luminous XRBs now known to exist. Models that explain these events
through \emph{bona fide} super-Eddington accretion \citep{begelman:2002}
should result in a different luminosity distribution and number of sources 
(per unit SFR) than models involving randomly-directed relativistic beaming
\citep{king:2001a}. Careful population synthesis can provide a means to
discriminate between these hypotheses, and we hope to report the result in
the near future.

Another relevant parameter concerns the magnitude of natal kicks to be
imparted to a neutron star or black hole at formation.  There are two
mechanisms for natal kicks. First, Blaauw-Boersma kicks
\citep{blaauw:1961} resulting from conservation of momentum after the
supernova which gives birth to the compact object. Second, asymmetric
kicks arise from the anisotropic emission of neutrinos and/or core
material in the supernova event. Compared to symmetric Blaauw-Boersma
kicks, a smaller amount of mass loss is needed to generate a
comparable velocity change, and this is especially true if the bulk of
the momentum is carried away in an anisotropic neutrino flux. For
neutron stars formed in a supernova, this scenario is
sufficient. However, if the star is massive enough to form a black
hole, there is a potential problem for the natal kick
scenario. \citet{gourgoulhon:1993} convincingly demonstrate that, if
the event horizon forms on the dynamical time scale of the collapsing
core, an insufficient number of neutrinos escape to drive a supernova
explosion through envelope heating. This implies a maximum mass for a
supernova progenitor, above which the supernova is quenched by the
event horizon before it begins. We have chosen a simple criterion for
whether a black hole will receive an asymmetric kick during collapse;
namely, all objects below 40~\Msun\ (referring to the ZAMS mass)
experience a random kick. Above this limit, objects collapse directly
to a black hole, with no kick. This is a simplification consistent
with hydrodynamical simulations such as those of \citet{janka:1996},
\citet{fryer:1999} and \citet{fryer:2000}.

The magnitude and direction of the asymmetric kick are still matters
of considerable debate. One strong possibility is that of
neutrino-induced convection, as discussed in
\citet{janka:1994,fryer:2000}, and references therein. In this
process, the angular momentum acts to stabilize the forming compact
object, so that rapidly-rotating progenitors produce
substantially-weakened explosions. The neutrino convection in
rapid-rotators is concentrated at the slowly-rotating poles, driving
an asymmetrical supernova. It is interesting that the kick vector
depends not only upon the rotation of the progenitor, but that an
inverse correlation is posited between the magnitude of the supernova
event and the asymmetry of the explosion. Unfortunately, the binary
evolution code does not track the rotation state of the progenitor,
and so while the above theoretical predictions are an interesting path
for future investigations, they cannot be effectively applied
here. Therefore, we take asymmetric kicks to be oriented isotropically. 
The imparted kick speed is selected from a Maxwellian distribution with an
energy corresponding to a 1.4 \Msun neutron star with a speed of 90
\kmsec. The distribution is truncated at the high end, with a maximum
kick speed of 500 \kmsec\ (again, for a 1.4 \Msun neutron star).  All
kick speeds are scaled to the mass of the recoiling object; e.~g., a 7
\Msun black hole will receive a speed change one-fifth the size that
would be imparted to the aforementioned neutron star. This kick is
slightly lower than that posited by \citet{han97}, and we do not
consider the possibility of a bifurcated kick distribution put forward
in recent literature \citep{arz:2002,pfahl:2002c}.

It is important to note that, in order for a system to become an XRB
of any type, it must first survive the natal kick produced when the
compact object is formed. The likelihood of this obviously drops
dramatically as the imparted kick speed increases. Since the kick
speed is inversely proportional to the mass of the compact object
progenitor, an immediate prediction is that XRBs with a black hole
accretor should be more common than the assumed IMF would indicate, as
these systems will survive the first natal kick more easily than
systems with a neutron star primary.  A similar effect holds for the
mass of the donor star, though in this case the effect is independent
of the kick magnitude. Systems which retain a larger fraction of their
total mass have a greater chance of surviving the first supernova;
this implies that binaries with more massive secondaries are also more
resistant to disruption. The overall effect is to increase the ratio
of HMXB to LMXB systems compared to what would be predicted on the
basis of the initial IMF alone.

\subsection{Implementation of Mass Transfer in the Code}

The code itself tracks changes of state, which means that instead of
evolving a binary system along a smoothly-flowing time axis, the next
evolutionary ``event'' (for example, a star may leave the main
sequence, or become a helium star after casting off its envelope
through a stellar wind) is found from the input stellar evolution
tracks, and the code advances the time index accordingly. At this
point the code extrapolates mass loss from winds for each star for the
elapsed time, and recalculates orbital parameters accordingly. Sudden
mass loss (from supernova events) is handled identically, with the
code reporting the evolutionary state and orbital parameters
immediately before the explosion, and immediately after. If, after
advancing to the next evolutionary state, the code determines that the
two stars should have interacted via mass transfer at some point, the
system is backed up to the immediately previous state, and the orbital
parameters are set such that (at least) one star is barely in contact
with its Roche-lobe. An episode of mass transfer is then resolved,
ending with both stars inside their respective Roche-lobes (or with a
spiral-in, if the system was undergoing common-envelope evolution and
lost sufficient orbital energy to bring the stellar cores into
contact, after ejecting the envelope). 

Mass transfer rates were computed following the model of
\citet{hurley:2002}, and references therein. To determine the
stability and time scale of mass transfer, adiabatic coefficients are
employed, which describe the response of the donor star's radius to
mass loss. These coefficients are defined in \citet{webbink:1985}, as
follows:
\begin{equation}
\zeta_{ad} \equiv\left(\frac{\partial\,\ln R}{\partial\,\ln M}\right)_{X,s}\; , \quad
\zeta_L \equiv \left(\frac{d\,\ln R_L}{d\,\ln M}\right) \; , \quad {\rm and} \quad
\zeta_{eq} \equiv \left(\frac{\partial\,\ln R_{eq}}{\partial\,\ln M}\right)_X\; ,
\end{equation}
where $\zeta_{ad}$ is a logarithmic derivative of the donor's radius
with respect to mass, at a constant chemical composition and specific
entropy, $\zeta_{L}$ is the logarithmic derivative of the donor's
Roche lobe radius with respect to its mass, and $\zeta_{eq}$ is the
logarithmic derivative of the radius of the donor in thermal
equilibrium, when held at a fixed chemical composition, with respect
to mass. The mode of accretion is determined from the following
inequalities between the coefficients

\begin{description}
\item[$\zeta_L < (\zeta_{ad}, \zeta_{eq})$] Nuclear time scale mass transfer.
Mass transfer is not self-sustaining, and is strongly dependent on the
degree to which the Roche lobe is overfilled. The resulting mass transfer
rate is
\begin{equation}
-\dot{M_2} = f(M_2) \;\ln\left(\frac{R_2}{R_L}\right)^3 
~\mbox{\rm M}_{\odot}~\mbox{\rm yr}^{-1}\; ,
\end{equation}
where $M_2$, $R_2$, and $R_L$ refer to the mass, radius and Roche lobe radius 
of the donor star, and $f(M_2)$ is given by
\begin{equation}
f(M_2) = 3\times10^{-6} \; \left[\min\left(M_2 , 5.0\right)\right]^2\; .
\end{equation}
This relation is chosen to ensure steady mass transfer \citep[and references
therein]{hurley:2002}.

\item[$\zeta_{eq} < \zeta_L < \zeta_{ad}$] Mass transfer occurs on the
thermal time scale of the donor's envelope. If $M$ is the donor mass, $M_c$
is the core mass of the donor, and $\tau_{kh}$ is the Kelvin-Helmholtz
time scale (in years) of the donor envelope, then the mass transfer rate is
\begin{equation}
-\dot{M_2} = \frac{M_2 - M_c}{\tau_{kh}} ~\mbox{M}_{\odot}~\mbox{yr}^{-1}\; .
\end{equation}

\item[$\zeta_{ad} < \zeta_L$] Dynamical mass transfer. In this situation,
the radius of the primary expands more quickly than the Roche surface after
transferring a mass element. Thus the mass-loss rate is limited only by the
sound speed in the envelope of the donor, and is a runaway process. If
$\tau_{dyn}$ is the sound-crossing time of the donor (the donor's radius 
divided by the envelope sound speed), then the mass-loss rate is just
\begin{equation}
-\dot{M_2} = \frac{M_2 - M_c}{\tau_{dyn}} ~\mbox{M}_{\odot}~\mbox{yr}^{-1}\; .
\end{equation}

\end{description}

A number of HMXBs exhibit a very different mode of mass transfer;
namely, accretion from a strong stellar wind coming off of a massive
companion, typically a Be- or O-star. We do not consider these systems
for a number of reasons. First, the X-ray luminosity of such an XRB is
highly variable, and strongly tied to the positions of the two stars
relative to the line of sight, as the absorption column density is far
from isotropic. Second, these systems tend to be very faint \citep[of
order $10^{36}~\ergs$ or less; see, for example,][]{yokogawa:2002},
unless in a rare outburst from an instability in the companion. While
these sources have a hard X-ray spectrum, their luminosities are very
low and they are never present in sufficient concentration to
dramatically alter the outcome of the simulation.

Pulsars and the associated supernova remnants (SNRs) also contribute
to the overall X-ray luminosity. While SNRs can attain luminosities of
$10^{37} \ergs$ and above \citep{martin:2002}, much of this radiation
is emitted below 2~keV, while we are interested in the hard X-ray
emission in the 2--10~keV band.  We also ignore isolated,
rotation-powered pulsars and their associated synchrotron nebulae
since their contribution is not significant unless the bulk of
newly-formed pulsars would require spin periods of 10~ms or less
\citep[see the discussion of][]{vanbever:2000}.

\subsection{Mass Accretion, Stability, and Resulting X-ray Luminosity\label{model:luminosity}}

For the Roche-lobe overflow systems we are considering here, the
accretion flow, collimated through the inner Lagrange point, naturally
forms into a disk structure around the accretor, as it cannot shed
angular momentum quickly enough to permit a direct impact onto the
primary \citep{shakura:1973}.  A distinction must be drawn between the
mass transfer rate (the rate of material passing through the inner
Lagrange point), and the mass accretion rate (the rate of material
accreting on to the compact object).  Clearly, the X-ray luminosity is
determined by the latter. In a conservative mass transfer scenario, no
matter is lost from the system, and the two transfer rates should be
equal if the disk remains in steady state. This is not generally true,
however, and we must quantify the extent to which mass transfer is
non-conservative. We can establish a relation defined in terms of the
Eddington luminosity limit of the accretor; so the mass transfer rate
$-\dot{M}_2$, and mass accretion rate $\dot{M}_1$, are related by
\begin{equation}
\dot{M}_1 = \left(\alpha_{Edd} - 1\right)\,\dot{M}_2\; ,
\end{equation}
where
\begin{equation}
\alpha_{Edd} = \max \left(0\;,\;1 - \frac{\dot{M}_{Edd}}{|\dot{M}_2|}\right)
\end{equation}
The $\dot{M}_{Edd}$ term is the mass accretion rate that generates a
luminosity equal to the Eddington luminosity of the accretor. In other
words, as the mass transfer rate climbs above $\dot{M}_{Edd}$, the
mass transfer becomes increasingly non-conservative, as $\alpha_{Edd}$
approaches unity.

The prescription for converting a mass transfer rate into a bolometric
luminosity follows from energy conservation. Assuming that the accreting
matter falls effectively from infinity onto the surface of a compact object
of radius $r_1$, the total available  power is
\begin{equation}
L_{bol} = \frac{G M_1 \dot{M}_1}{r_1}\; .
\label{eqn:Lbol}
\end{equation}
Half of the potential energy liberated by the infalling material is
released in the accretion disk before making contact with the compact
object \citep[e.g.,][]{pringle:1985}. Precisely where this energy is
emitted becomes important when considering accretion onto a black
hole, where energy crossing the event horizon cannot be radiated.
Thus, we introduce the efficiency parameter $\eta_{bol}$, with a value
of 0.5 for black hole accretors, representing the fraction of energy
released before the accreting material disappears across the event
horizon. For neutron star accretion, we take this value to be unity,
with essentially all of the energy radiated from either the inner
accretion disk and boundary layer or from the accretion column, in the
case of a strongly magnetized object.

A second efficiency factor, $\eta_{st}$, that must be considered is
the duty cycle of the accreting systems. In particular, the dramatic
high-luminosity, soft-spectrum and low-luminosity, hard-spectrum
states observed in Galactic HMXBs such as Cygnus~X-1 and GS~1124--68
can be understood as a transition between a bright, optically-thick
disk and a dim, radiatively-inefficient accretion flow
\citep{meyer:2000}. Such transitions are thought to be the result of
an instability analogous to the ``dwarf nova'' instability seen in
many cataclysmic variables. A thorough analysis of the physics that
drives the instability can be found in \citet[p. 104]{frank:1992} and
\citet{hameury:1990}.  Since we cannot compute the structure and
stability properties of the accretion disk, we adopt an empirical
approach based on observational information. It is known that LMXB
systems with black hole accretors (BH/LMXBs), for example, almost
always exhibit the dwarf nova instability. For neutron-star LMXBs
(NS/LMXBs) this is much less common.  It is thought that significant
X-ray emission from accretion on to the neutron star illuminates and
heats the disk, maintaining the high-viscosity state and allowing
stable accretion \citep{vanparadijs:1996,king:1996,king:2001}.

For the purposes of our simulation, we have broken down this problem into
three possible cases, as follows:

\begin{enumerate}

\item
{\it Accretion onto a neutron star primary from a massive companion
(high- or medium-mass X-ray binaries; a.k.a. NS/HMXB or
NS/MMXB). --} This is the simplest case. The generally high mass
transfer rates from massive companions, coupled with the radiation
from the surface of the neutron star, act to prevent the dwarf nova
instability. We therefore take the duty cycle of such systems to be
unity.

\item
{\it Accretion onto a neutron star from a low-mass companion
(NS/LMXB). --} Here, the lower mass transfer rate means less
illumination of the accretion disk by accretion onto the neutron star.
To determine if the instability manifests itself in a given system of
this type, we employ a period criterion inferred from figure 3
in \citet{li:1998}.  The
critical period $P_{crit}$ is
\begin{equation}
\log P_{crit}\; {\rm (days)} = 6.5 \left(\frac{M_2 - 0.6}{\Msun}\right)\; ,
\end{equation}
where $M_2$ is the donor mass, as before. If the donor mass is below
0.6~\Msun, the energy released by mass transfer is judged to be
insufficient to stabilize the disk, regardless of the orbital
separation of the system. If $0.6\; \Msun < M_2 < 0.8\; \Msun$, then the
system is taken to be stable if the orbital period is less than the
critical period $P_{crit}$ above. If $M_2$ exceeds $0.8 \Msun$, then
the system is stable only if its period is less than twenty days. All
systems with larger periods are assumed to exhibit the dwarf nova
instability, and are assigned a duty cycle of 0.1, meaning that each
system has a 10\% chance of being counted in the overall X-ray
luminosity. This value is on the high end of a typical duty cycle for soft
X-ray transients \citep{tanaka:1996}.

\item
{\it Black hole accretors from a companion of any mass. --} Because
there is little radiation from the vicinity of the compact object
illuminating the disk, a black hole system must be closely bound to
avoid the disk instability. The criterion for this case is taken from
\citet{king:2001}, where it is derived from estimating when the
luminosity required to suppress disk outbursts exceeds the Eddington
luminosity. The critical period for the system in this case is
\begin{equation}
P_{crit} \simeq 3.3 \left(\frac{f_{disk}}{0.7}\right)^{-1.5} 
\left(\frac{\dot{M}_2}{0.5\; \dot{M}_{Edd}}\right)^{0.75}
\left(\frac{M_1}{M_2}\right)^{0.125}\quad\mbox{days}
\end{equation}
The $f_{disk}$ term refers to the disk filling fraction, which is the
ratio of the disk radius to the radius of the accretor's Roche
lobe. We estimate this radius as the point at which the disk is
disrupted by the tidal forces exerted by the donor star. Those systems
with periods above $P_{crit}$ are flagged as subject to the dwarf nova
instability. The duty cycle of such systems is known to be
dramatically shorter than for neutron star accretors with this disk
instability. This is because, in the absence of illumination from
accretion onto a stellar surface, more material must be added to the
disk to raise the temperature to the point where steady accretion can
occur, and this material takes more time to accumulate. Typical
recurrence times are on the order of a decade or longer \citep{chen:1997}
,implying duty cycles
of 0.01 or lower. With such extended recurrence times, it is difficult
to estimate a mean time between outbursts, as few transient systems
have been observed in outburst more than twice. For now, we assume a
duty cycle of 0.01, keeping in mind that this will likely change as
more black hole transient systems are catalogued.

\end{enumerate}

A final efficiency parameter, $\eta_x$, is employed to convert the
calculated bolometric luminosity to the power radiated between
2--10~keV. This factor varies with the type and spin of the accreting
compact object. For simplicity (and because the code does not track
spin), we assume that 40\% of the bolometric luminosity from accretion
onto a black hole is emitted in this band. For neutron star accretion,
that value is 20\%. This adjustment is used everywhere that an X-ray
luminosity is quoted herein.

To summarize, the integrated 2--10~kev luminosity of the populations
is computed as
\begin{equation}
L_{\rm 2-10\; keV} = \sum_{\rm all\; XRBs} 
\eta_{st}\; \eta_x\; \eta_{bol}\; 
L_{bol}\; ,
\label{eqn:Lx}
\end{equation}
where the last term represents the total available accretion power as
given by equation~(\ref{eqn:Lbol}) and all the other terms are the
efficiency factors discussed above. The first efficiency factor,
$\eta_{st}$, which represents the duty cycle of accretion-disk
instabilities, is a true multiplicative factor only in the ideal case
of an infinitely large population of XRBs. In practice, however, we
are dealing with a finite number of systems and the effect of
instabilities is a stochastic fluctuation of the X-ray luminosity, as
individual systems experience high and low states. This behavior is
particularly obvious at late times when only a small number of systems
contribute to the observed X-ray luminosity.  In order to capture this
stochastic behavior, we have adopted a monte-carlo approach to
assigning values to $\eta_{st}$. More specifically, for each system we
assign a value of either 0 or 1 to $\eta_{st}$ with a probability
determined by the instability duty cycles noted above. This correction
is also applied to the number of ``active'' systems presented with our
results in the following section.

\section{Simulation Results\label{model:results}}

Four simulation runs were performed, each simulating a 20 Myr episode
of star formation, at a constant rate of 10~\Msunyr. Of the four
simulations, two were performed using the Salpeter initial mass
function \citep{salpeter:1955}, and two with the Miller-Scalo IMF
\citep{miller:1979}. For each IMF, one run was performed with a flat
mass ratio distribution, $\psi(q) = constant$, and the other draws the
mass ratio from the distribution $\psi(q) = 2/(1 + q)^2$, which
approaches the spectroscopically-determined distribution shown in
equation~(\ref{eqn:q}). For ease of reference these shall be referred
to hereafter as ``Sal/flat~$q$'', ``Sal/low~$q$'', ``MS/flat~$q$'',
and ``MS/low~$q$''.

The first question to address is the normalization of the simulations;
that is, how one converts from a desired star formation rate and
duration to a total number of binaries generated. Since the code draws
the primary mass from an IMF that is truncated at 4~\Msun\ at the low
end, we need a weighting factor, $w$, which is the number of generated
stars of all masses divided by the number of generated stars above the
4~\Msun\ cutoff. In other words,
\begin{equation}
w = \frac{\int_{0.1}^{\infty} dN}{\int_4^{\infty} dN}
\end{equation}
where $dN$ is the differential number of stars with mass $m=M/{\rm
M}_{\odot}$, i.~e., the IMF. Performing this integration we obtain
$w=145.5$ for the Salpeter IMF. Similarly, by considering the three
separate power laws of the Miller-Scalo IMF, we obtain $w = 69.2$ for
this case.

Next, a definition of the binarity fraction is needed. We take the binarity
fraction $b$ to represent the fraction of \emph{systems} that contain two
stars. For example, if a sample of three stars exists as a binary pair and
one single star, the binarity is taken to be one-half (as opposed to
two-thirds).

If we let $\bar{m}$ and $\bar{q}$ represent the average primary mass
(in \Msun) and binary mass ratio, respectively, and let $n$ denote the
total number of binaries formed per unit time (in a 1 Myr period),
then the total SFR for the 1 Myr interval is the sum of three parts;
the mass contributed by the primary stars, the mass contributed by the
companion stars (weighted by $\bar{q}$), and the mass of the single
stars which are not considered in the code but are generated according
to the binarity $b$. Assuming a binarity of $b = 0.5$, we can then
write
\begin{equation}
\mbox{SFR}
=  nw\bar{m} + nw\bar{m}\bar{q} + nw\bar{m}\left(\frac{1 - b}{b}\right) 
=  nw\bar{m} (2 + \bar{q})
\end{equation}
We note that $\bar{m}$ and $\bar{q}$ are functions of the IMF and mass
ratio distribution, respectively, while $w$ is a function of the IMF.
The values of each of these parameters for the four runs are shown in
Table~\ref{table:params}. The last column of the table shows the
number of binary systems generated for each Myr of the simulation,
representing a constant 10 \Msunyr SFR. The SFR varies linearly with
the number of binaries formed in each interval, allowing the results
to be adjusted for an arbitrary SFR by simple proportionality. This
proportionality breaks down at large times, however, when large
scatter is introduced by small-number statistics, as only one or two
sources are active at any given moment.

\begin{figure}
\epsscale{1.2}
\plotone{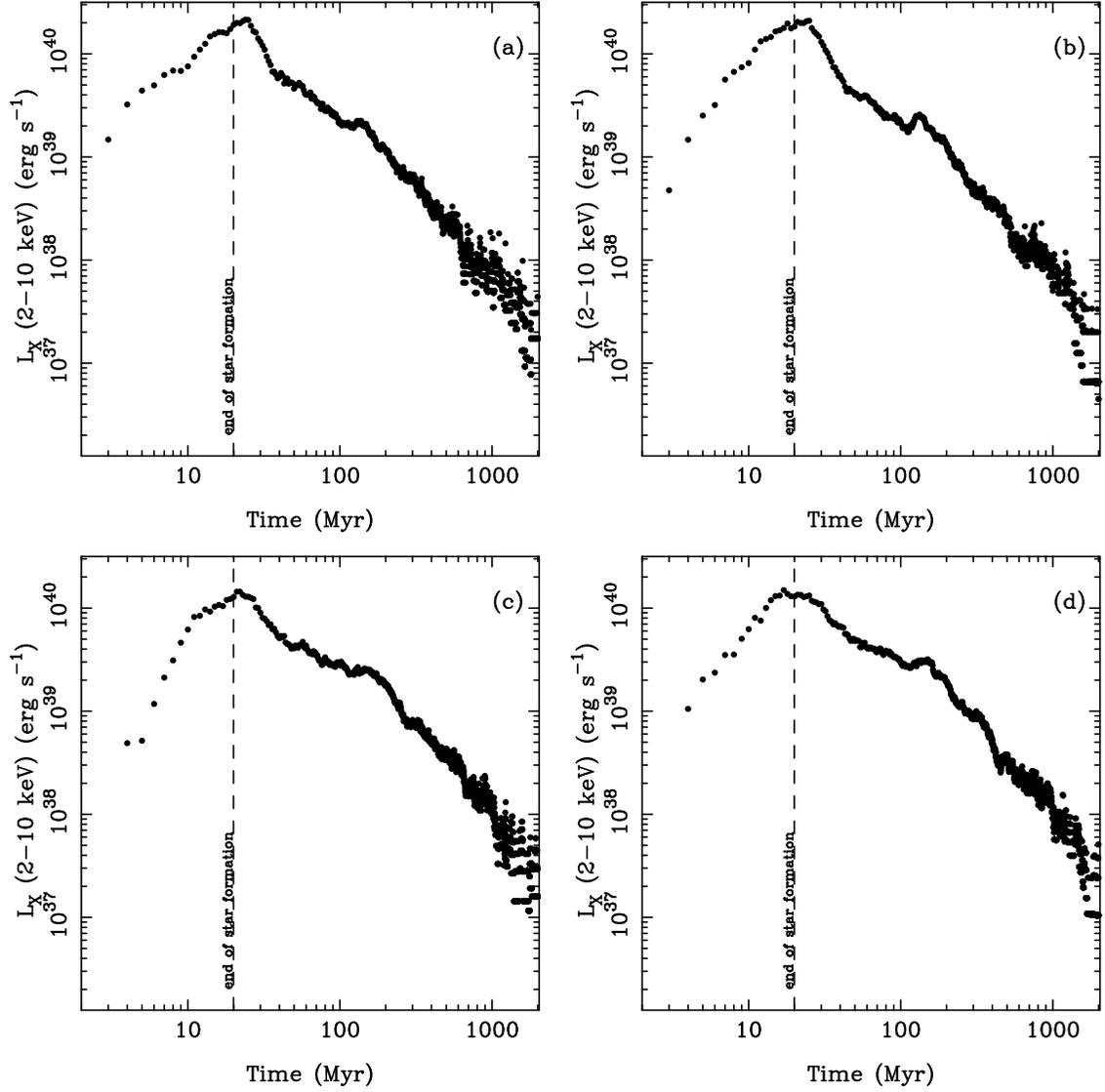}
\caption[Evolution of total X-ray luminosity]{\small The evolution of the
  total 2--10~keV X-ray luminosity of a simulated population over 2
  Gyr, from a SFR of 10~\Msunyr, extending for 20~Myr
  (vertical dashed line). The luminosity includes all efficiency factors,
  following equation~(\ref{eqn:Lx}). Panel (a) shows the results of a Salpeter
  IMF with a low-skewed $q$-distribution. Panel (b) also uses the
  Salpeter IMF, with a flat mass ratio distribution. Panels (c) and
  (d) use the Miller-Scalo IMF with a low-skewed and flat mass ratio
  distribution, respectively.
\label{fig:lx}
}
\end{figure}

\begin{figure}
\epsscale{0.8}
\plotone{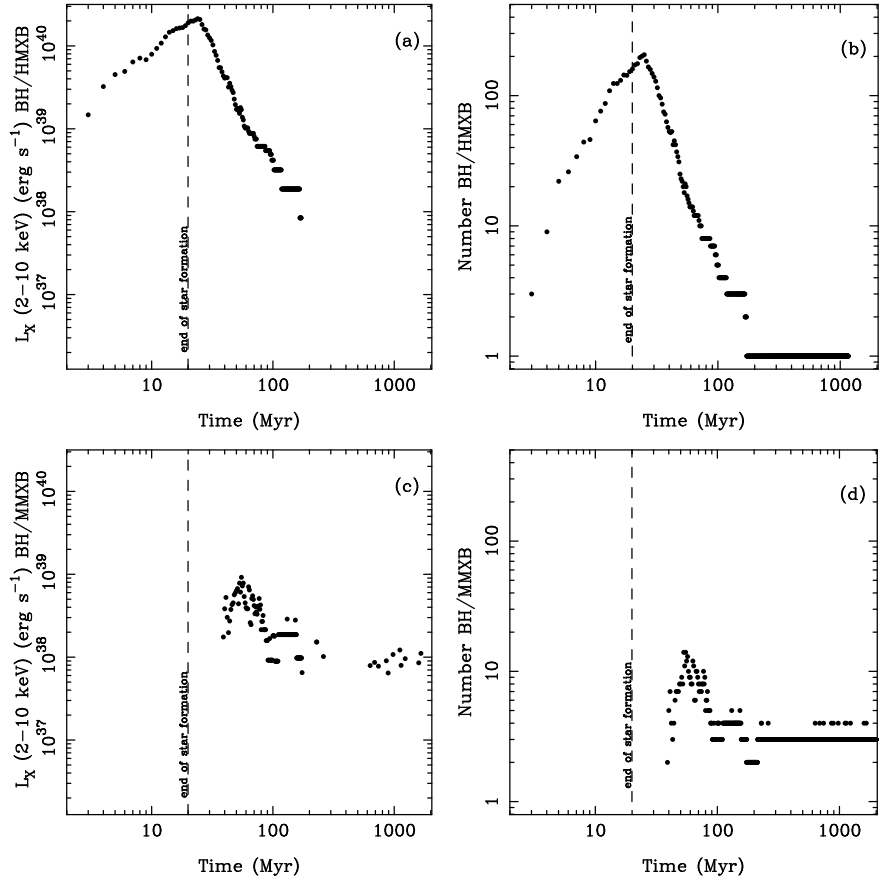}
\caption[Evolution of BH/XRB population, Salpeter IMF, low q]{\small 
  The evolution of the 2--10~keV luminosity of the BH/XRB component of the
  population over 2~Gyr, from a SFR of 10~\Msunyr,
  extending for 20~Myr (vertical dashed line), for a Salpeter IMF and
  a low-skewed q-distribution.  The luminosity plots include all
  efficiency factors, following equation~(\ref{eqn:Lx}). The population
  size plots show the number of {\it active systems} only (i.~e.,
  systems in a low state are excluded). Note the short delay between
  the start of the simulation and maximum luminosity, roughly
  corresponding to the nuclear lifetime of the secondary (donor)
  star. The luminosity evolution for (a) the BH/HMXB, and (c) the
  BH/MMXB populations are shown, alongside the evolution track of each
  population's size (b,d). No BH/LMXB population was detected in the
  first 2 Gyr epoch.
\label{fig:sal_lowq_bh}
}
\end{figure}

\begin{figure}
\epsscale{0.8}
\plotone{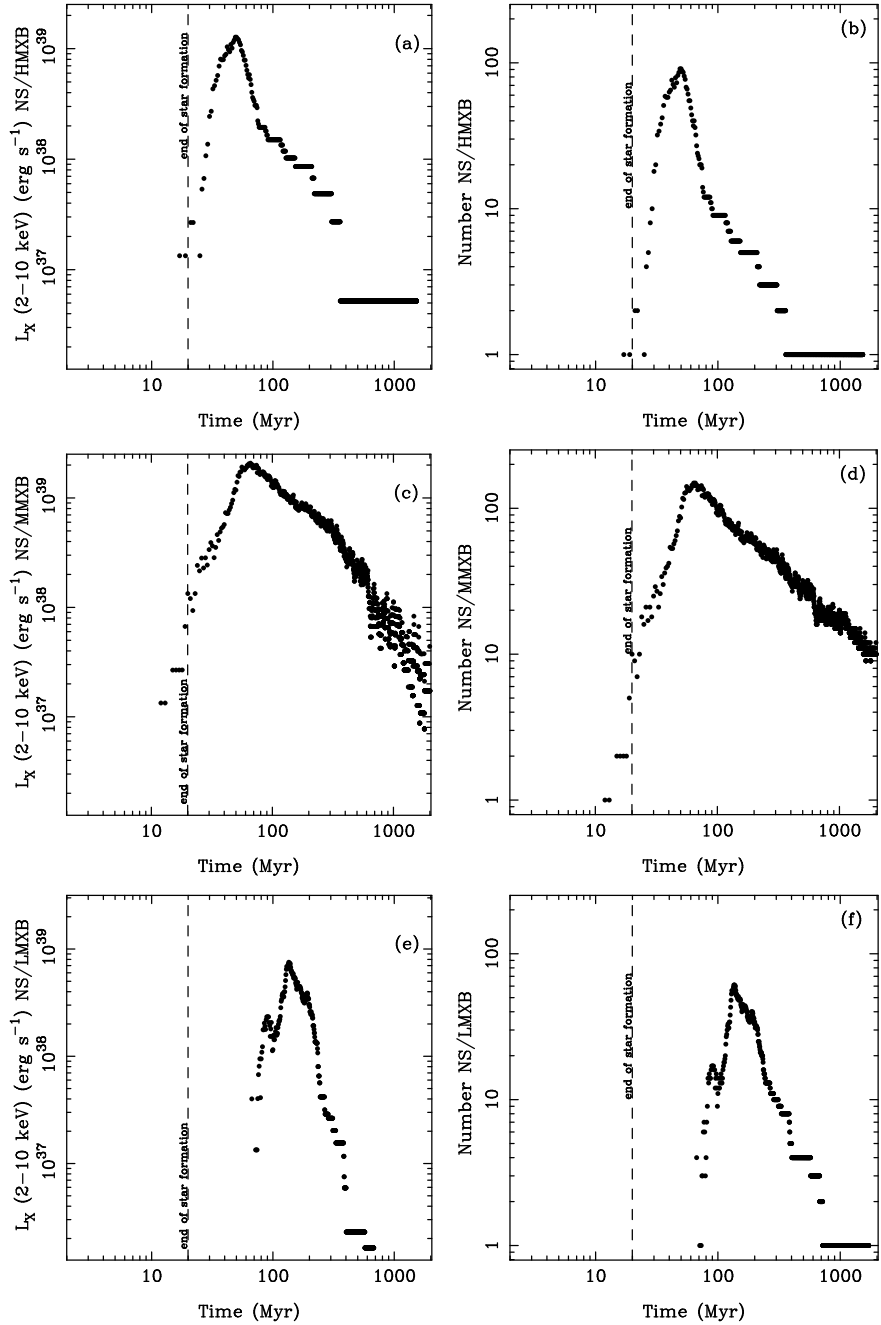}
\caption[Evolution of NS/XRB population, Salpeter IMF, low q]{\small The
  evolution of the 2--10~keV luminosity and size of NS/XRB component
  of the population over 2 Gyr, from a SFR of
  10~\Msunyr, extending for 20~Myr, and for a Salpeter IMF with a low-skewed
  q-distribution.
  The luminosity plots include all
  efficiency factors, following equation~(\ref{eqn:Lx}). The population
  size plots show the number of {\it active systems} only (i.~e.,
  systems in a low state are excluded). This is the analogue of
  Figure~\ref{fig:sal_lowq_bh} but for binaries with neutron star
  primaries.  The luminosity evolution for (a) the NS/HMXB, (c) the
  NS/MMXB, and (e) the BH/LMXB populations are shown, alongside the
  evolution track of each population's size (b,d,f).
\label{fig:sal_lowq_ns}
}
\end{figure}

\begin{figure}
\epsscale{0.8}
\plotone{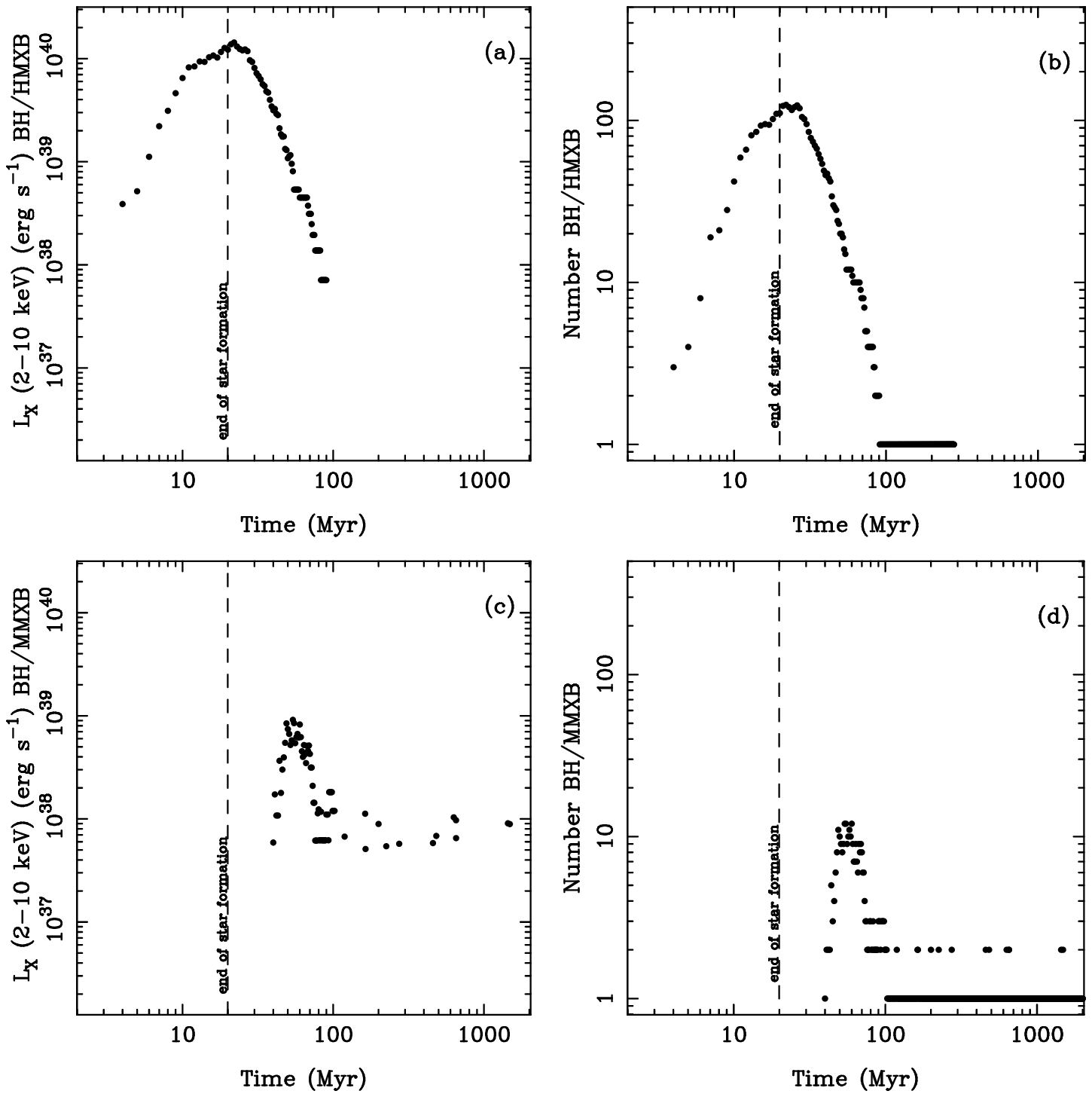}
\caption[Evolution of BH/XRB population, Miller-Scalo IMF, low q]{\small
  Same as Figure~\ref{fig:sal_lowq_bh}, for a Miller-Scalo IMF and a
  low $q$ distribution.
\label{fig:ms_lowq_bh}
}
\end{figure}

\begin{figure}
\epsscale{0.8}
\plotone{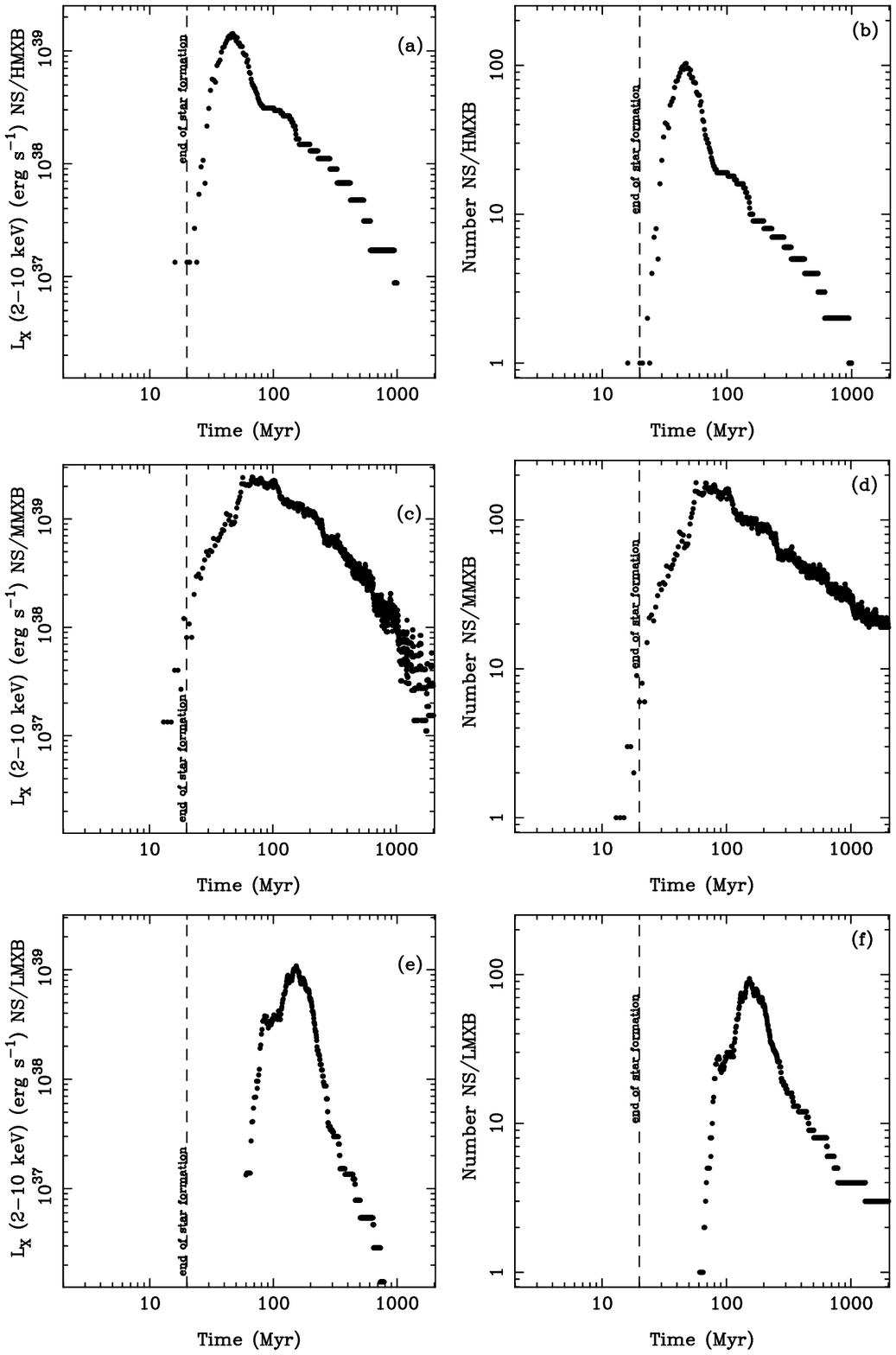}
\caption[Evolution of NS/XRB population, Miller-Scalo IMF, low q]{\small
  Same as Figure~\ref{fig:sal_lowq_ns}, for a Miller-Scalo IMF and a
  low $q$ distribution.
\label{fig:ms_lowq_ns}
}
\end{figure}

\begin{figure}
\epsscale{0.8}
\plotone{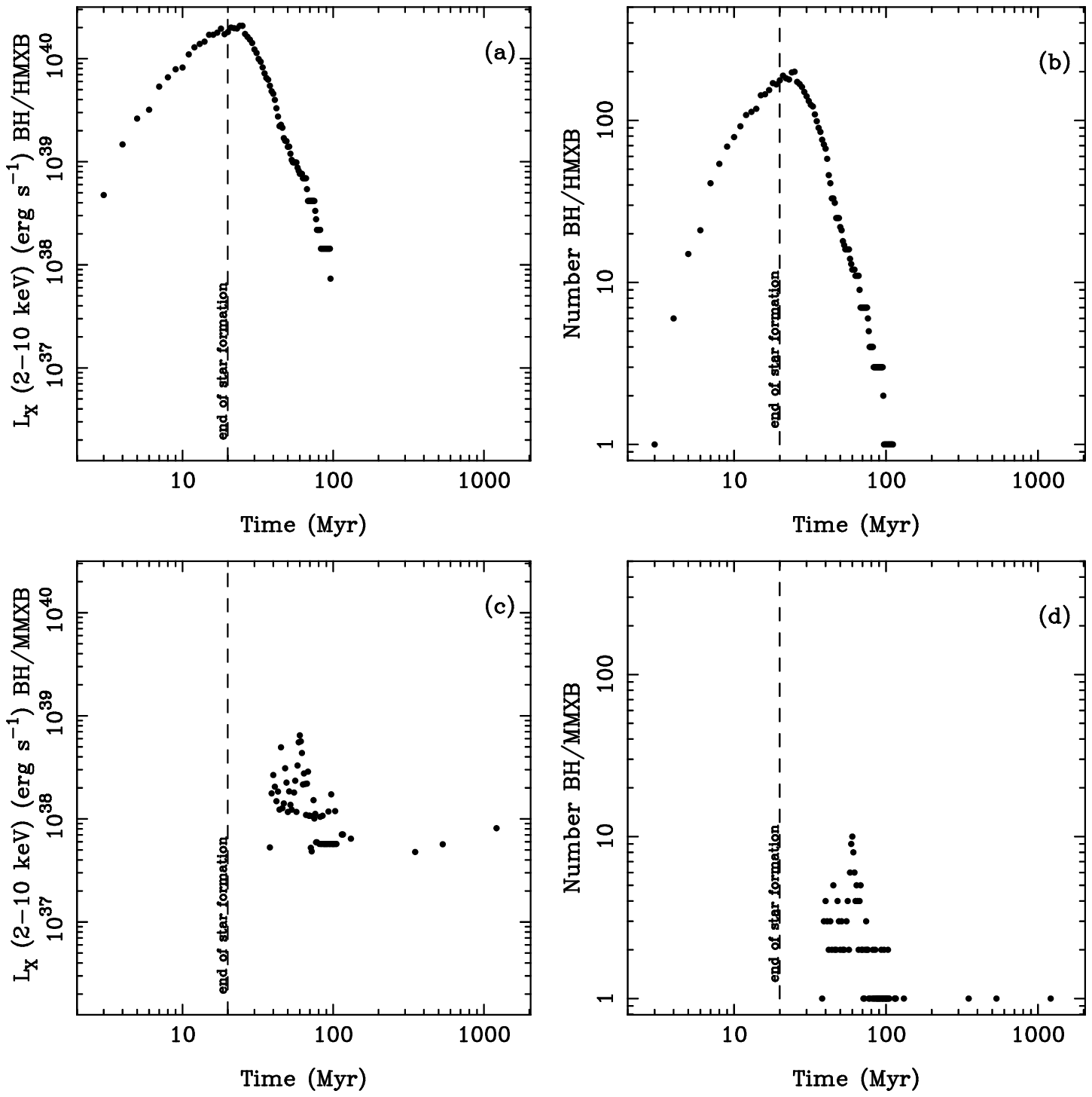}
\caption[Evolution of BH/XRB population, Salpeter IMF, flat q]{\small Same
  as Figure~\ref{fig:sal_lowq_bh}, for a Salpeter IMF and a flat $q$
  distribution (all values of $q$ equally likely).
\label{fig:sal_flatq_bh}
}
\end{figure}

\begin{figure}
\epsscale{0.8}
\plotone{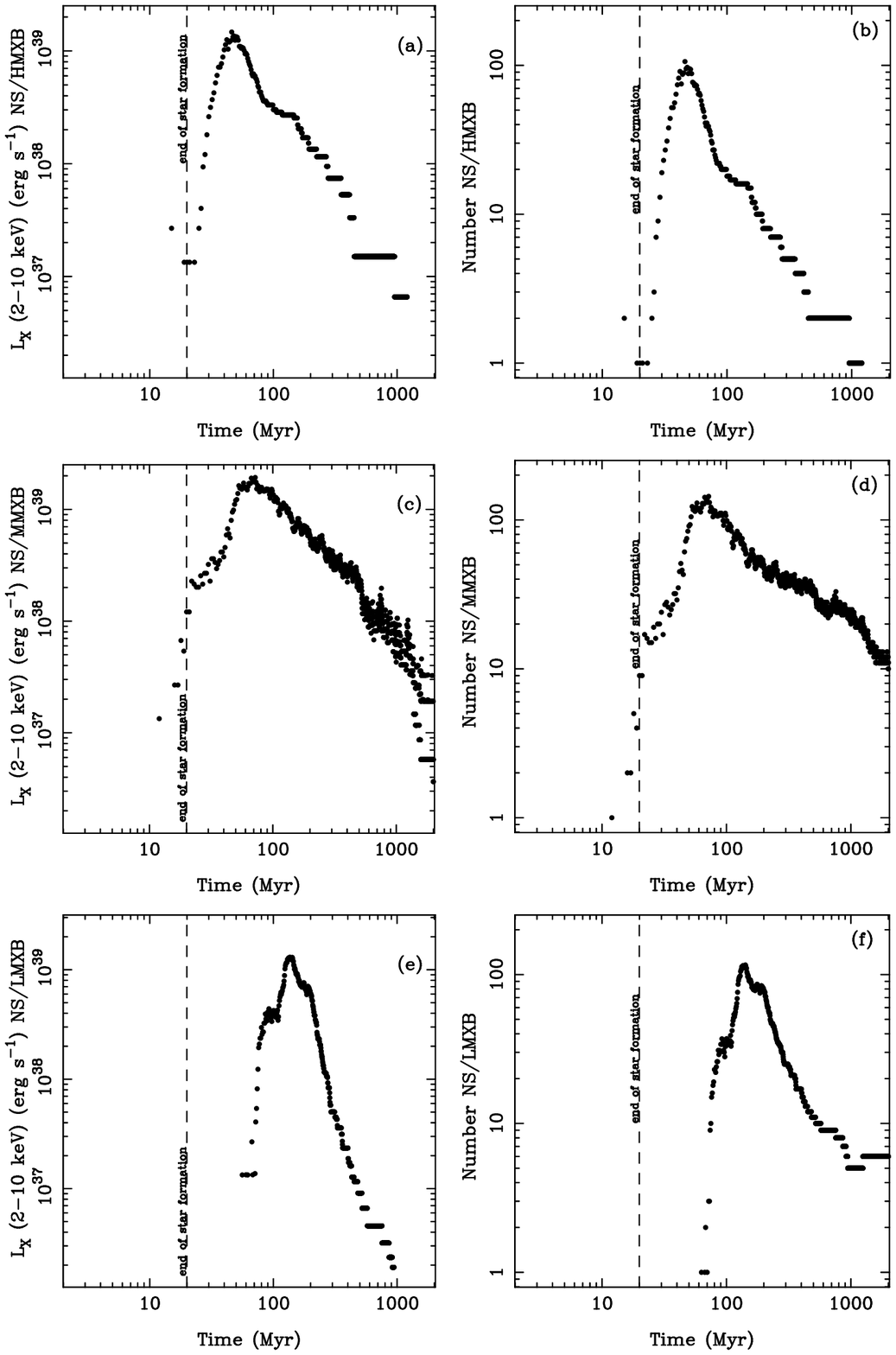}
\caption[Evolution of NS/XRB population, Salpeter IMF, flat q]{\small Same
  as Figure~\ref{fig:sal_lowq_ns}, for a Salpeter IMF and a flat $q$
  distribution (all values of $q$ equally likely).
\label{fig:sal_flatq_ns}
}
\end{figure}

\begin{figure}
\epsscale{0.8}
\plotone{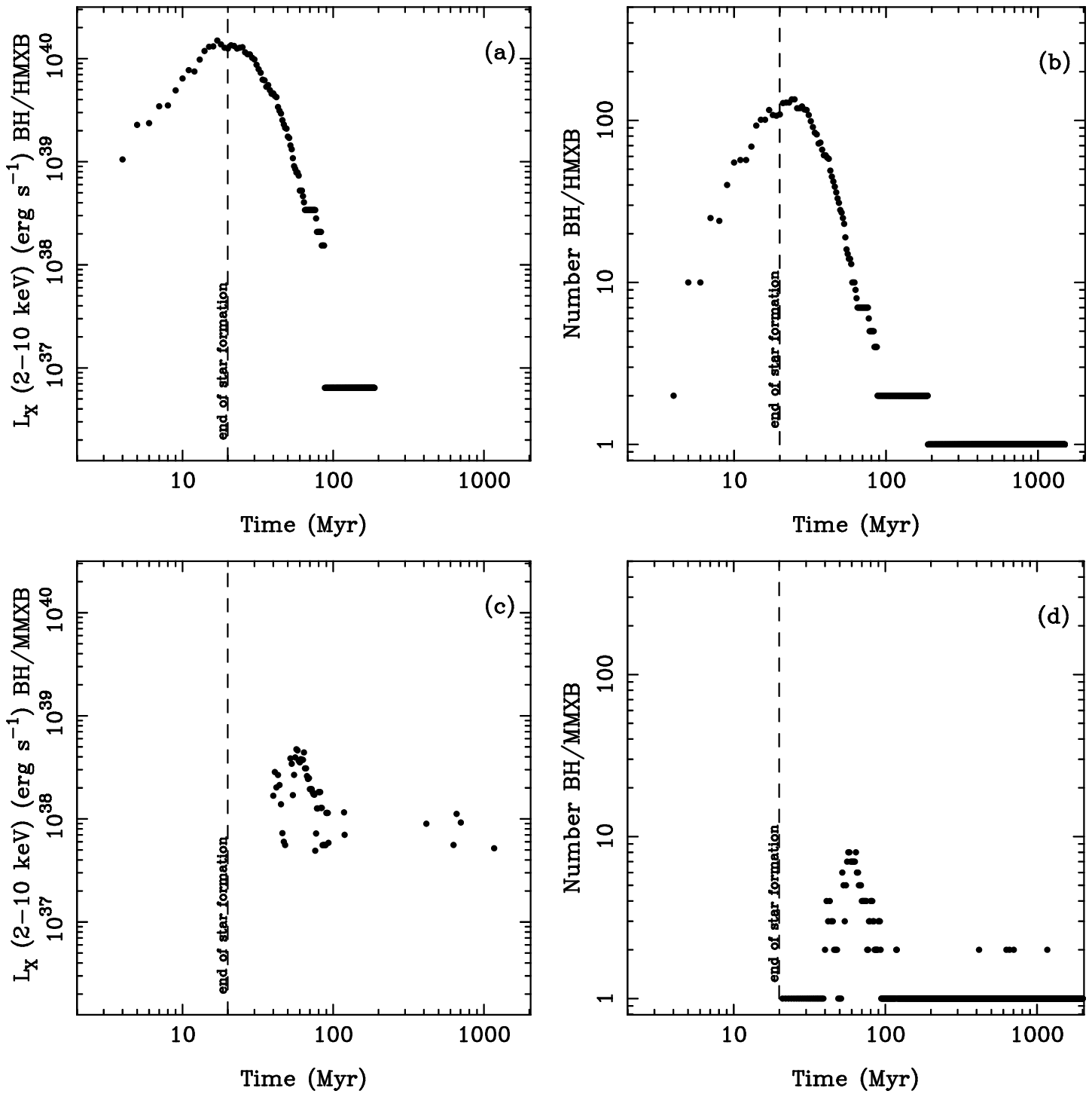}
\caption[Evolution of BH/XRB population, Miller-Scalo IMF, flat q]{\small
Same as Figure~\ref{fig:sal_lowq_bh}, for a Miller-Scalo IMF and
a flat $q$ distribution.
\label{fig:ms_flatq_bh}
}
\end{figure}

\begin{figure}
\epsscale{0.8}
\plotone{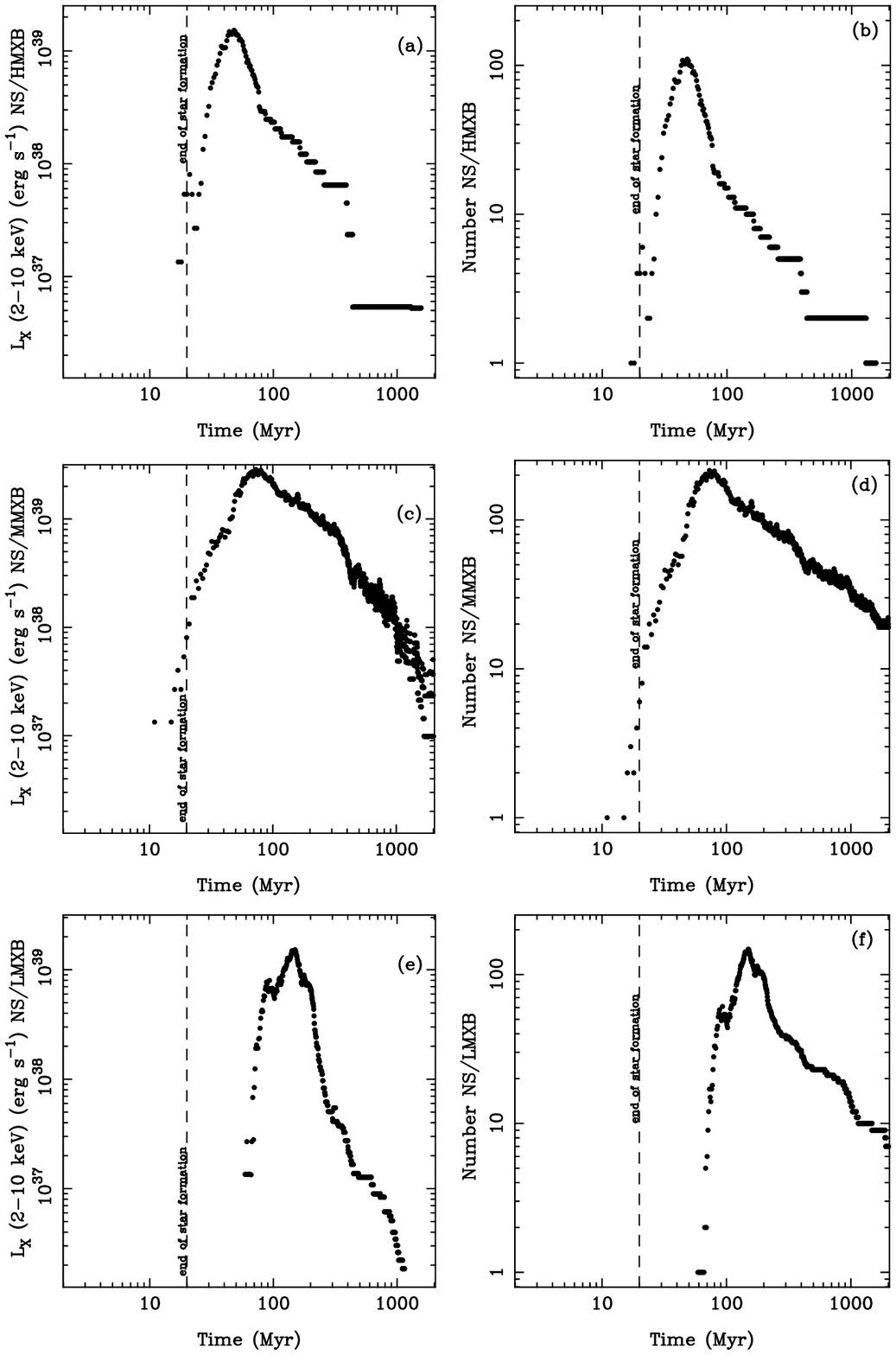}
\caption[Evolution of NS/XRB population, Miller-Scalo IMF, flat q]{\small
  Same as Figure~\ref{fig:sal_lowq_ns}, for a Miller-Scalo IMF and a
  flat $q$ distribution.
\label{fig:ms_flatq_ns}
}
\end{figure}

\begin{figure}
\epsscale{0.8}
\plotone{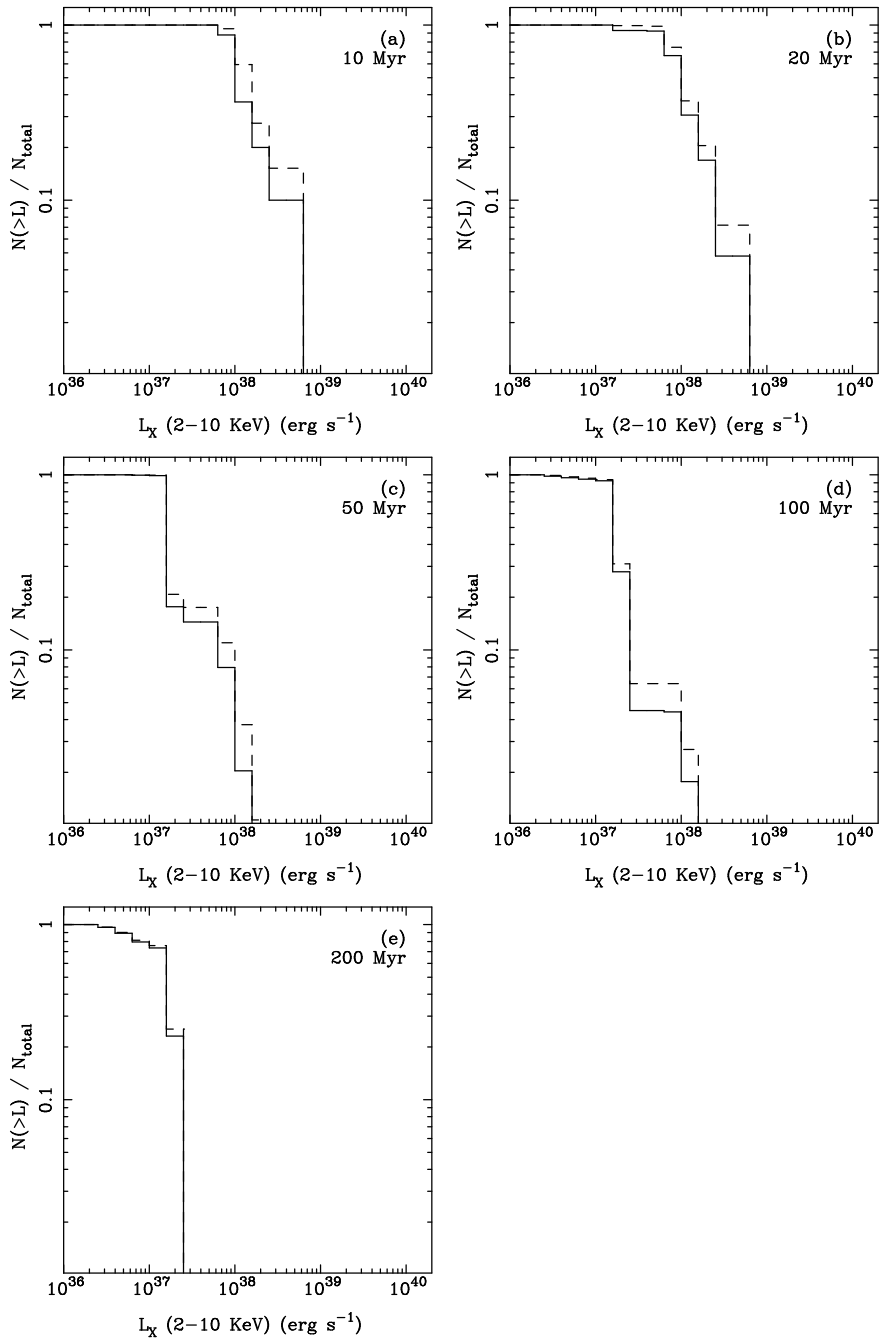}
\caption[Luminosity function at five epochs, Salpeter IMF, low q]{\small The
  cumulative luminosity function for the population derived from the
  Salpeter IMF, and a low-skewed $q$ distribution. The five epochs are
  (a) 10 Myr, (b) 20 Myr, (c) 50 Myr, (d) 100 Myr, and (e), 200
  Myr. These were chosen to show the dramatic change as star formation
  ends and massive stars no longer dominate (a and b), as the X-ray
  luminosity becomes driven by systems with increasingly lighter
  secondaries (c, d and e). The solid an dashed lines show the upper and
  lower bounds of this function, resulting from shirt-term stochastic 
  variations in the number of active systems (see \S\ref{model:luminosity}
  and \S\ref{model:results} for details).

\label{fig:sal_lowq_lf}
}
\end{figure}

\begin{figure}
\epsscale{0.8}
\plotone{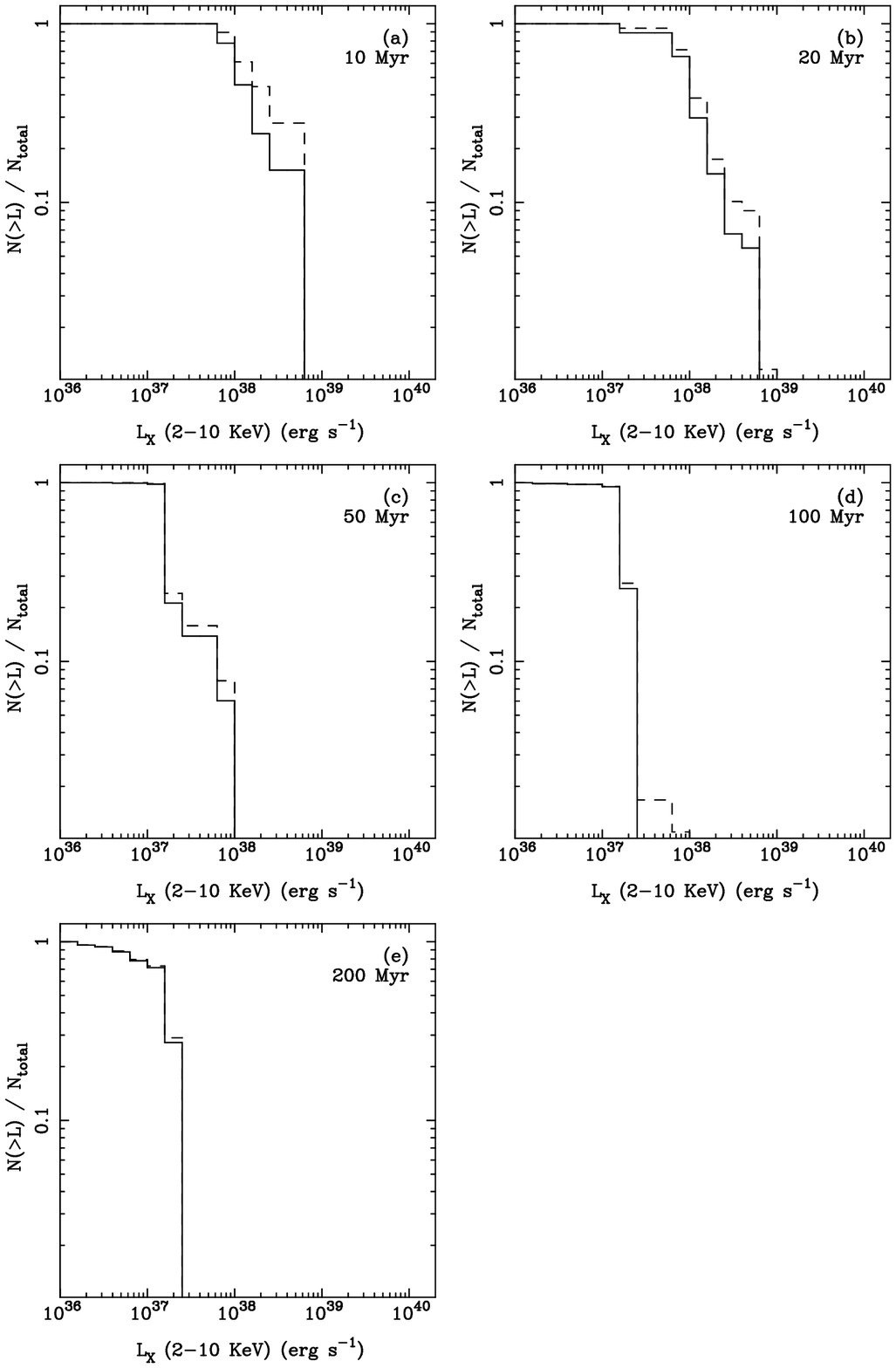}
\caption[Luminosity function at five epochs, Miller-Scalo IMF, low q]{
Same as for Figure~\ref{fig:sal_lowq_lf}, for a Miller-Scalo IMF and 
a low $q$ distribution.
\label{fig:ms_lowq_lf}
}
\end{figure}

\begin{figure}
\epsscale{0.8}
\plotone{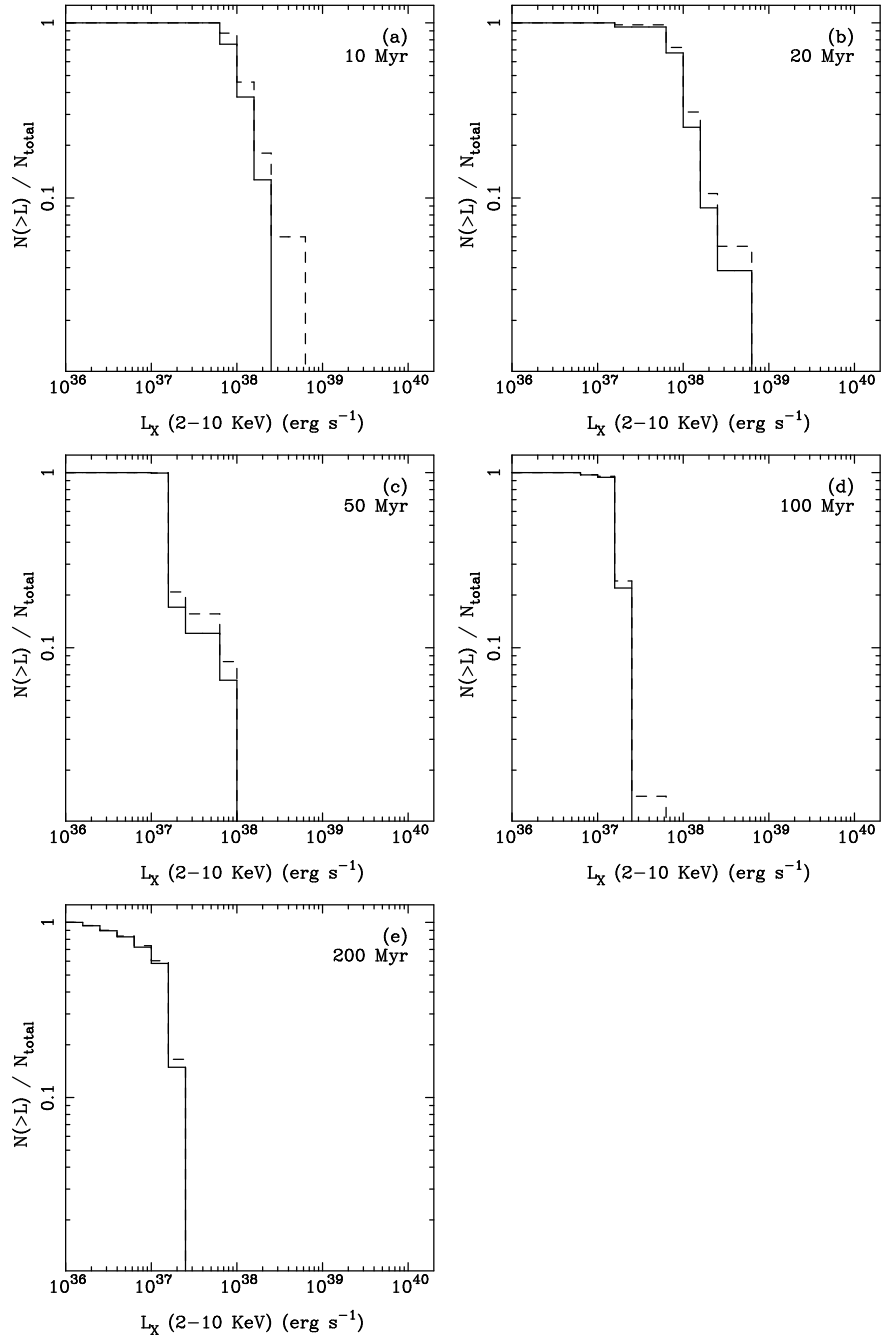}
\caption[Luminosity function at five epochs, Salpeter IMF, flat q]{\small
  Same as for Figure~\ref{fig:sal_lowq_lf}, with a Salpeter IMF and a
  flat $q$ distribution.
\label{fig:sal_flatq_lf}
}
\end{figure}

\begin{figure}
\epsscale{0.8}
\plotone{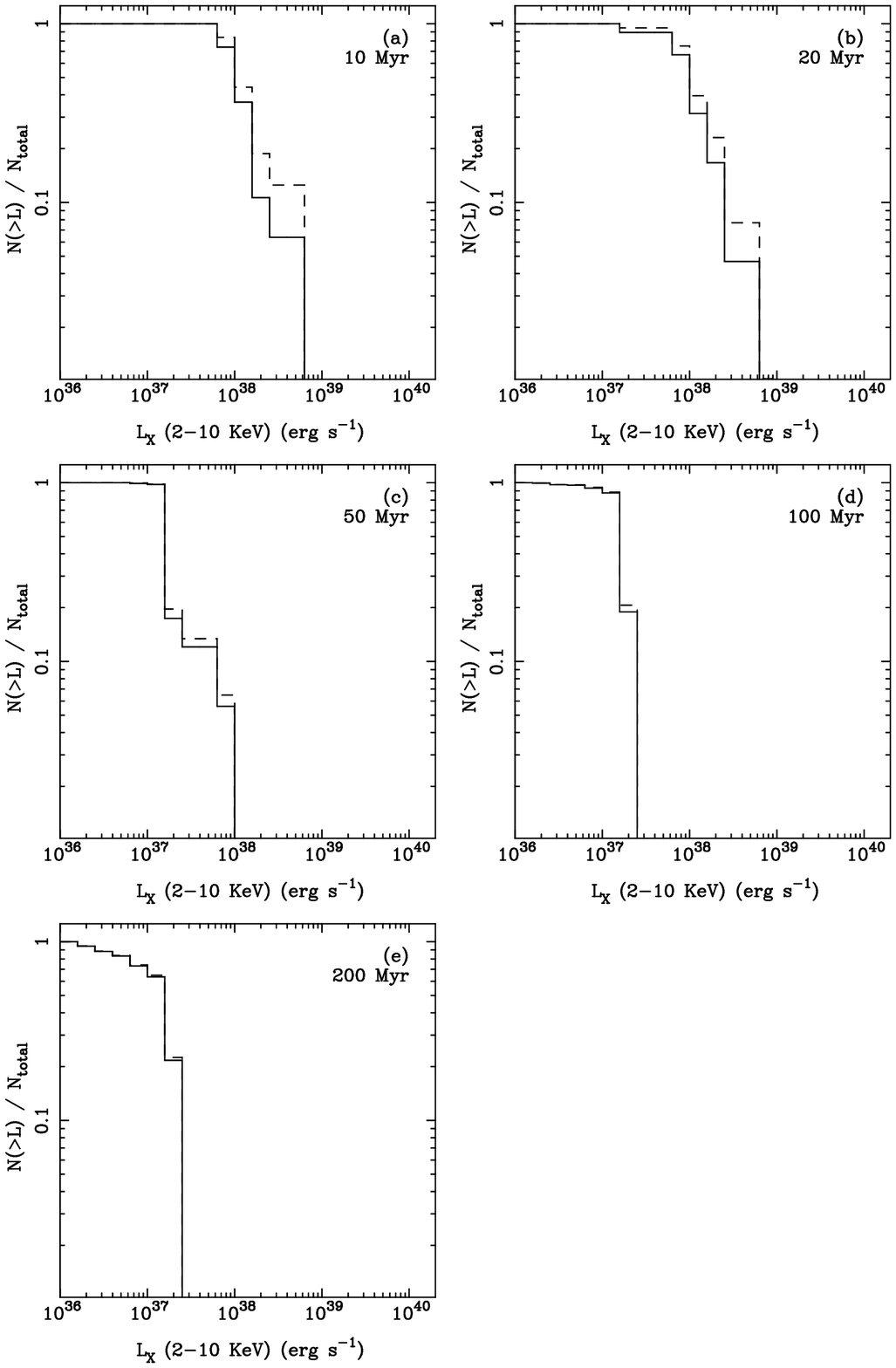}
\caption[Luminosity function at five epochs, Miller-Scalo IMF, flat
  q]{\small Same as for Figure~\ref{fig:sal_lowq_lf}, with Miller-Scalo IMF,
  and a flat $q$ distribution.
\label{fig:ms_flatq_lf}
}
\end{figure}

One of the principal results of the simulations are the curves showing
the time evolution of the 2--10~keV X-ray luminosity (calculated
according to \S\ref{model:luminosity}) and population size of the
X-ray luminous population (with stochastic corrections for high/low
states applied).  The population is subdivided into six categories,
based on two intrinsic properties of the system. First, the accretor
is classified as either a black hole or a neutron star. Second, the
donor is placed in one of three mass categories. Donors with a mass
less than 1.4~\Msun\ are LMXBs, donors with a mass between $1.4$ and
8~\Msun\ (the minimum mass necessary to form a neutron star) are
considered to be medium-mass X-ray binaries (MMXBs), and more massive
donors are, naturally, HMXBs. Plots showing the evolution of the
luminosity and size of each population subset are shown, one set for
each of the four possible parameter sets.
Figures~\ref{fig:lx}--\ref{fig:ms_flatq_lf} detail the final output of
the investigation.

Figure~\ref{fig:lx} gives the most important and directly observable
result; namely, the total 2--10~keV X-ray luminosity for the
population, evolved over the first 2 Gyr after the star formation
episode. The luminosity includes all efficiency factors, following
equation~(\ref{eqn:Lx}). Panels (a) and (b) show the results for a
Salpeter IMF, with a low-skewed and flat $q$-distribution,
respectively, while panels (c) and (d) show the analogous plots for a
Miller-Scalo IMF.  It is apparent from Figure~\ref{fig:lx} that the
input IMF and mass ratio distribution have little or no effect on the
X-ray luminosity evolution of the population. This is an interesting
and important result, because it implies that the output depends
little upon uncertainties in the IMF, or choice of $q$-distribution,
both parameters with significant uncertainties. Conversely, however,
this also means that it would be difficult to constrain these
parameters through a comparison of the results with observational
data.

Figure~\ref{fig:sal_lowq_bh} shows the evolution of XRBs with black
hole accretors, given an initial Salpeter IMF, and a low $q$
distribution after equation~(\ref{eqn:q}). Each point represents the
luminosity or population size for a 1 Myr interval. Star formation
progresses at a constant rate for 20 Myr; the end of star formation is
shown as a vertical dashed line on each of the plots.  The left two
panels (a and c) show the luminosity evolution of the HMXB and MMXB
populations, respectively. No active BH/LMXBs were found in the first
2 Gyr epoch. The corresponding plots on the right side show the number
of {\it luminous} X-ray emitting systems of each type at the indicated
epoch (i.~e., systems in their low states are excluded). The evolution
of all systems was tracked out to a maximum of 2 Gyr, though massive
systems evolve on more rapid time scales, and are only tracked until
the second compact object is formed. Panel (a) shows the luminosity
evolution for the BH/HMXB systems, with a peak at about
$2\times10^{40}\,\ergs$, coinciding with the end of the star formation
episode. This is understood to be the result of massive donor stars,
with short main sequence lifetimes. HMXBs thus form very quickly after
the creation of the initial binary, as the secondary (donor) star
leaves the main sequence. The HMXB population tracks star formation
closely, with accreting systems accumulating until star formation
stops, at which point the remaining systems quickly die off; 20 Myr
after the end of the star formation episode, no HMXBs remain. One
final point: comparing the peak luminosity in panel (a) with the peak
BH/HMXB population size in panel (b) of the same figure, we see that
the mean luminosity of a binary is around $10^{38}$ \ergs. This
corresponds to the Eddington luminosity limit for an accretor of
roughly 5~\Msun\ (with all efficiency factors taken into account),
which the typical black hole mass given the assumed IMF and the
progenitor-to-black-hole mass mapping, discussed above. Thus, we
conclude that the black holes, which dominated the hard X-ray luminosity
at early times, accrete very close to their Eddington limit.

Continuing with Figure~\ref{fig:sal_lowq_bh}, panels (c) and (d)
detail the evolution of BH/MMXB population, with companion masses
ranging from 1.4--8~\Msun. The first and most important thing to
notice is the delay in the onset of this population. The first such
systems do not appear until more than 10 Myr have elapsed, and their
numbers do not peak until about 30 Myr after the end of star
formation. This is a result of the longer evolutionary time scales of
these less-massive companion stars. The black hole forms quickly, but
the binary will only enter the accretion stage on the nuclear
time scale of the donor star. This concentration of sources and
corresponding peak in X-ray luminosity at $\sim50$--$60$ Myr is a
result of two effects. First, because of the choice of mass ratio
distribution, a primary massive enough to form a neutron star (at
least 8~\Msun) is less likely to have a light companion. For a flat
mass ratio distribution, the average companion mass for an 8~\Msun\
star is 4~\Msun. For the low-skewed mass ratio distribution, this
value drops to just under 3~\Msun. This is a selection effect of
sorts, arising from the minimum mass necessary to form a neutron star.
The second factor to consider is that the duration of a typical mass
transfer episode occupies a larger fraction of the life of a massive
star than a star at the light end of the MMXB category ($\sim1.4~{\rm
M}_{\odot}$). Since massive companions are formed on a time scale
comparable to their main sequence lifetimes, they all tend to leave
the main sequence and fill their Roche lobes in a similarly short
period. This contributes to the luminosity peak in panel
(c). Longer-lived stars are less likely to leave the main sequence in
closely-timed groups, spreading out the resulting X-ray binaries in
time. These sources are not concentrated in time, but XRBs from this
group continue to ``turn on'' even at very large times (several Gyr).

One other feature of note in panel (c) is the bifurcation in the
luminosity evolution that occurs about 70 Myr into the
simulation. This population oscillates rapidly between a
low-luminosity and a high-luminosity state. A glance at the number
evolution plot shows that, at this epoch, less than ten sources (and
at times as few as two) comprise the entire MMXB group. The variation
is therefore partly due to small number statistics, as single sources
flicker in and out of mass transfer episodes. As well, in the systems
comprising the upper branch of the bifurcation the companion star is
crossing the Hertzprung Gap, implying a rapid expansion and
commensurately high mass transfer rate and luminosity. The lower
systems tend to be transferring matter on the nuclear time scale of the
companion, with a lower resulting luminosity per system.

Figure~\ref{fig:sal_lowq_ns} is analogous to
Figure~\ref{fig:sal_lowq_bh}, and shows the evolution of populations
in which the compact object is a neutron star rather than a black
hole. All run parameters remain identical. Panels (a) and (b) are
almost precisely the same as those of the previous figure, with the
exception of a noticeable delay in the rise-time of the NS/HMXB
population.  This is directly related to the longer main sequence
lifetime of the lighter neutron star progenitors.

Panels (c) and (d) show the evolution of the NS/MMXB set. This is the
most numerous population, and, at its peak value is the most luminous
as well, for a brief period from roughly 40--80~Myr into the
simulation. The evolution curves are quite smooth, until around
200~Myr, when the XRB population size fluctuates rapidly between
several and ten or twenty sources at a time, with a corresponding
dramatic fluctuation in the luminosity. The slow decrease in the
maximum luminosity as the population ages results from fewer accreting
systems, and also lower mass transfer rates from increasingly lighter,
longer-lived companion stars. It is interesting to note, however, that
the luminosity of this subset of XRBs remains significant well beyond
even 2 Gyr. This means that populations of NS/MMXBs formed in
long-vanished starbursts continue to contribute to the X-ray
luminosity of their host galaxy, long after the galaxy return to
quiescent star formation. This also means that successive waves of
rapid star formation may allow NS/MMXBs and NS/LMXBs to accumulate, so
that the total number of such systems observed in the present era
represents a number of individual starbursts taking place in the
distant past.

Panels (e) and (f) show the NS/LMXBs. Note that they also attain a
significant luminosity, at a later epoch than the MMXB population.
One might expect that no such systems would be present in the first 2
Gyr of the simulation, as the main sequence lifetime of even the most
massive companions in this category (1.4~\Msun) is significantly
longer than that.  These systems are all survivors of an episode of
common envelope evolution, during which the primary star ascended the
giant branch, and enveloped the secondary.  The small orbital
separation required for the companion to fill its Roche-lobe is
unlikely to have occurred \emph{ab initio} given the distribution of
initial orbital separations.  The dramatic reduction in orbital
separation during this phase meant that, after the primary went
through a supernova, the secondary was close enough to begin
transferring mass onto the nascent neutron star. The small probability
of this outcome explains the relatively small number of systems in
this category.

The relationship between the high-mass and lower-mass XRB systems,
could best be summarized as follows. The BH/HMXB and NS/HMXB
populations track the \emph{current}, or \emph{recent} star formation
rate.  MMXBs and LMXBs, however, better track the
\emph{time-integrated star formation} in a galaxy's history, i.~e.,
the total mass of stars formed.

A quantity that is easily observable and can also be computed from the
results of our simulations is the luminosity function. If one assumes
that the luminosity function is a power law of the form
\begin{equation}
N(L) \equiv N_0\; L^{-\gamma} \quad ({\rm with~\gamma > 0}),
\end{equation}
the {\it cumulative} luminosity function is
\begin{equation}
N(>L) \equiv \int_L^{\infty} N(L)\; dL = N_1\; L^{-\beta}
\quad ({\rm for~\gamma\neq 1}),
\end{equation}
where $\beta = \gamma-1$ and $N_1=N_0/(1-\gamma)$. The cumulative
luminosity function is the quantity most often derived from
observations.  With this in mind, we show in
Figures~\ref{fig:sal_lowq_lf}--\ref{fig:ms_flatq_lf} the cumulative
luminosity function at five epochs (10 Myr, 20 Myr, 50 Myr, 100 Myr
and 200 Myr) after the beginning of star formation. The luminosity
function is expressed in terms of the fractional amount of the
population that exists above a given luminosity. One interesting
aspect of these functions is that, because they represent a time-slice
of the luminosity evolution, longer-lived systems are more likely to
be included. This makes them representative, but the high-luminosity
sources, which are only active for a short time, may be missed. This
means that the high end of all of the displayed luminosity functions
is subject to a considerable amount of variability as short-lived,
high-luminosity systems flicker on and off. To circumvent this, we
consider at each epoch 100 time slices, separated by $10^4$ years. For
each of these slices, we calculate a luminosity function, and then
plot the extremes of these functions as solid and dashed lines.

Table~\ref{tbl:lfs} shows the measured power-law indices of the cumulative
luminosity functions of Figure~\ref{fig:sal_lowq_lf} and those immediately
following. The ranges given correspond to the extremes of the luminosity
function at each epoch, as discussed in the previous paragraph.
As can be seen from this table, the principal determining factor for
this quantity is the elapsed time since the end of star formation.
This is because the luminosity function becomes dramatically steeper as
the population ages and luminous X-ray sources die off. In principle,
one could use this value to estimate the age of a stellar population. In
practice, this would require a great deal of detailed information about
the star formation history of the host galaxy, in order to segregate
individual stars by the particular star-formation episode which spawned
them.  We will return to the cumulative luminosity function and other
observables in \S\ref{model:observations}, to compare our predictions
with the results of recent observations.

\section{Discussion of Simulation Results}

An important question is the overall robustness of the results with
respect to several other variables that are not explicitly listed as
input variables, but comprise assumptions made in the implementation
of the population synthesis code. We now describe each of these in
detail, along with probable effects on the output, and a justification
for the choice of parameter value employed.

As mentioned previously, our choice for mass loss through winds during
the main sequence phase is taken from the work of
\citet{langer:1989}. This is the canonical treatment of winds used in
the majority of the literature. A more recent formulation by
\citet{hamann:1998} dramatically lowers the wind-loss rate on the main
sequence, and is used in the numerical simulations of
\citet{vanbever:2000}. This has the consequence of keeping the binary
separation smaller throughout the main sequence evolution (as a larger
mass loss would widen the system dramatically). As well, stars retain
a larger fraction of mass prior to the supernova, resulting in larger
compact object masses. The end result is closer, more massive systems
that are more likely to survive the first supernova without becoming
unbound. This results in a greater number of potential XRBs, and a
correspondingly larger X-ray luminosity. As an example, the results of
\citet{vanbever:2000} show an X-ray luminosity several times greater
than in our own simulations, resulting from a factor of four decrease
in wind mass loss.  The wind prescription of \citet{hamann:1998} has
only been applied to a few systems. Until confirmed by further
observations, the formulation of \citet{langer:1989} is still a very
reasonable choice for describing winds.

Magnetic breaking of the secondary star was taken into account via the
model of \citet{eggleton:2001}, which relates the torque to the
properties of the star, assuming that the secondary is in synchronous
rotation with the orbit of the binary. Although the adopted
prescription is relatively simple, magnetic breaking is an important
effect only in the case of stars with convective envelopes, i.e., the
secondaries of LMXBs. Since LMXBs make only a small contribution to 
the integrated X-ray luminosity, the precise prescription for magnetic 
breaking has a minimal effect on our results.

The mapping between the mass of a black hole progenitor and the final
mass of the hole (the black hole ``IMF'') is another
poorly-constrained quantity.  We make use of the IMF derived by
\citet{fryer:1999}, using hydrodynamic collapse models. Critical to
the end result is the mass at which a progenitor will collapse
directly to a black hole, with no intervening supernova (and hence no
asymmetric kick, discussed below). This limit strongly influences the
average resulting black hole mass, as objects above the limit tend to
retain most or all of their pre-collapse mass, and objects below this
limit lose a substantial fraction of this mass through a supernova. If
the critical collapse limit is high, most black holes will experience
a supernova at formation, with a lower resulting mass. If this limit
set low, the average black hole mass will be much greater, with
correspondingly narrower orbits and larger Eddington accretion
rates. The results in \citet{fryer:1999} do approximate the typical
observed mass in black hole candidates (averaging around
$\sim6\mbox{--}7\,\Msun$; \citealt{bailyn:1998}). However,
this observable is likely influenced by selection effects dependent on
the mass of the black hole.

The effect of kick magnitude on the survival of a binary is discussed
in Sipior \& Sigurdsson (2002).  The kick distribution chosen for our
simulations is a single Gaussian consistent with the bimodal
distribution of of \citet{podsiadlowski:2002b}, where a mix of low
($\sim90$~\kmsec) and strong ($\sim450$~\kmsec) kicks is shown to
reproduce the distribution of observed neutron star velocities.

The Eddington limit is a somewhat contentious restriction on the
maximum accretion rate of a compact object. In our models, we assume
that the Eddington rate applies weakly; that is, the mass accretion
rate is throttled back dramatically above this limit, with no abrupt
cutoff. Still, this only permits a minor violation (typically of a few
tens of percent) of the Eddington limit. However, there is
observational evidence of systems that exceed this boundary by factors
of several. For example, three X-ray pulsars in the Magellanic Clouds
the are known to exhibit substantially super-Eddington luminosities
\citep[namely LMC~X-4, SMC~X-1, and A~0538--66;][and references
therein]{nagase:1989,woo:1995}. Nevertheless, most galactic XRBs are
at or below this limit, so this is representative of most known
sources.

Lastly, we consider the parameter governing the efficiency of
common-envelope evolution. \citet{webbink:1984} introduced the idea of
of simulating a phase of common-envelope evolution by decreasing the
orbital energy by the binding energy of the envelope. The efficiency
parameter was assumed to be unity, so that the two quantities were
equal. Later work by \citet{dekool:1990} showed that when compared
against reasonable stellar models, the average efficiency parameter
dropped to around 0.5, and this is the value we use. This value means
that the orbit must lose twice the binding energy of the envelope in
order to dissipate it. It is true, however, that the precise value of
the efficiency parameter depends strongly on the structure of the
evolving star.  More recently, \citet{dewi:2000} showed that the
efficiency value ranged between 0.2 and 0.8 for the vast majority of
stellar evolution tracks.  Unfortunately, the code we are using does
not track the structure of the star closely enough for us to be able
to incorporate this more accurate prescription.  Nevertheless, it is
clear that the mean parameter value is a good choice overall for
calculating the end results of a common-envelope evolution
phase. Increasing the parameter would result in wider systems, and
hence fewer XRBs. However, the fact that fewer systems would merge
under these conditions would mitigate this somewhat. A lower value of
the efficiency parameter would dramatically increase the number of
mergers, with fewer systems surviving this phase to become XRBs.

\section{Comparison With Other Theoretical Work\label{model:discussion}}

Recently, a number of attempts have been made by various groups to use
population synthesis techniques to estimate the X-ray luminosity of a star
formation episode at various epochs. We discuss three of them here, and draw
a comparison between the results and methodology.

\subsection{Numerical Simulations by Van Bever \& Vanbeveren}

The simulations of \citet{vanbever:2000} are nearest to our own work
in terms of technique. A sizable population of binary systems was
generated, using a library of stellar evolution calculations detailed
in \citet{vanbeveren:1998a,vanbeveren:1998b,vanbeveren:1998c}. Winds,
in particular, receive a great deal of attention, as the precise mass
loss formalism used can greatly affect the final evolutionary outcome
of massive stars. In particular, the choice of wind strength will
change the mass distribution of black holes seen in the population, as
the black hole progenitor loses a different amount of mass prior to
collapse. As mentioned above, our simulations make use of the wind
mass loss formalism of \citet{langer:1989}. \citet{vanbever:2000} make
use of the mass loss rates of \citet{hamann:1998}, which are roughly
four times smaller than those predicted by \citeauthor{langer:1989},
and include the effects of line blanketing and clumping in determining
wind strengths. The result of these stronger winds are wider binaries,
but with a larger fraction surviving the initial supernova. More black
hole mass measurements are needed to lend support to the low wind mass
loss rates; in particular, a large population of black holes with
confirmed masses above $\sim10~{\rm M}_{\odot}$ would require winds
considerably weaker than those predicted by \citet{langer:1989}.

Another significant difference in the assumptions made by
\citet{vanbever:2000} involves the final collapse of a massive star to
a black hole. They assume that all such objects (those with an initial
mass of above 25~\Msun) collapse directly to a black hole, with no
associated supernova event. This again has the effect of increasing
the mean black hole mass, as no material can be lost from the system
via a supernova.  They themselves note that there is observational
evidence contravening this assumption; specifically, the dramatic
overabundance of O, S, Mg and Si in the atmosphere of the optical
component of the LMXB GRO~J1655--40, reported by
\citet{israelian:1999}. The primary in this system is a strong black
hole candidate, with an inferred mass of $6 \pm 2 {\rm
M}_{\odot}$. Our own simulations assume that a star with a zero age
main sequence (ZAMS) mass of 20~\Msun\ or more will become a black
hole. Those progenitors with a ZAMS mass of less than 40~\Msun\ are
assumed to undergo a supernova upon black hole formation. Above this
limit, the hole forms via direct collapse, with no explosion. The
difference is important, because black holes that form through direct
collapse should experience no natal kicks, which can act to disrupt
the system. The result is that a greater fraction of systems with a
black hole survive to become X-ray binaries. For the same reason, a
greater fraction of the observed XRB population will comprise
accreting black holes, since neutron stars will still experience kicks
at the same rate as before.

\citet{vanbever:2000} generate a single burst of $3 \times 10^5$
stars, selecting the masses of single stars and binary primaries from
a Salpeter IMF. A flat $q$ distribution is used to choose the mass for
the companion star. Single stars are tracked in this model because the
hard (2--10~keV) X-ray contribution from supernova remnants (SNRs) is
included in the total. To do this, neutron stars are assigned a
magnetic field strength and initial spin period chosen from a random
distribution, which allows the rotational energy loss rate to be
calculated. A small fraction ($\sim0.03$) is then used to determine
the amount of this energy that emerges as X-rays. We do not include
young supernova remnants in our own calculations, for two
reasons. First, the assignment of a magnetic field and rotational
period at formation ignores the importance of the evolutionary history
of the progenitor star, and so is fairly arbitrary. Second, the
contribution of remnants to the total X-ray luminosity should be quite
small for all but the shortest period pulsars. Indeed, after
completing their simulation, \citet{vanbever:2000} come to very much
the same conclusion, claiming that SNRs contribute to the total
starburst X-ray luminosity only when most or all neutron stars are
born with an initial period of less than 10~ms.

\citet{vanbever:2000} plot the X-ray luminosity evolution for the
first 10 Myr after the starburst. The onset time for the X-ray
luminous phase is 3--4 Myr, precisely the same as in our own
simulations. The peak {\it specific} luminosity of around $10^{33}
\ergs~{\rm M}^{-1}_{\odot}$ is reached shortly thereafter (around 5
Myr), and remains constant through the 10 Myr that are plotted. This
is very different from our own results, which show that the X-ray
luminosity continues to rise nearly 100 Myr after the end of star
formation, though the rate of increase slows dramatically after star
formation ends. By only tracking the population out to 10 Myr,
\citet{vanbever:2000} missed the important contribution of neutron
star accretors to the overall X-ray luminosity over an extended period
of time.  Note that the specific luminosity is given in terms of a
power per unit solar mass of stars generated. If we scale the
luminosity results shown in Figure~\ref{fig:lx} by dividing the peak
luminosity by the total mass of stars formed in each simulation ($2
\times 10^8~{\rm M}_{\odot}$), we obtain a specific luminosity of
approximately $5 \times 10^{32}~\ergs~{\rm M}^{-1}_{\odot}$, almost
exactly the peak rate given by \citet{vanbever:2000}.

One last thing to note about both the results of \citet{vanbever:2000}
and our own simulations is the strongly stochastic behavior of the
luminosity evolution. This is noticeable throughout the 10 Myr range
considered by the former, and can be seen in our results at longer
time scales (typically above 100 Myr). This is the direct result of
small-number statistics, when few systems are in an active state at a
given time. The \citet{vanbever:2000} results are based on a relatively
small initial population of $3\times10^5$ stars, many of which will not
become XRBs. Our larger population size of $2\times10^6$, and the fact
that the star formation occurs over a significant interval, helps to
smooth out dramatic fluctuations in the luminosity until a much larger
period of time has elapsed.

\subsection{Analytic Calculation by Wu}

As an alternative to the population modeling shown heretofore, where
individual systems are tracked from birth to evolutionary end state,
\citet{wu:2001} views a population of XRBs in terms of differential
equations governing birth and death rates. Unfortunately, to formulate
such a system of equations requires a slew of simplifying assumptions. In
particular, \citeauthor{wu:2001} assumes perfectly circular initial orbits
to calculate gravitational radiation and magnetic braking time scales,
and completely ignores the effects of natal kicks from supernovae. In
addition, no allowance is made for disk instability effects. As well, the
parameterization of many input quantities make comparison with observables
difficult, with no clear coupling to a SFR. To keep the
system integrable analytically, the initial population and birth rate
of binaries are taken as somewhat ad hoc power-laws in luminosity. Also,
the lifetime for X-ray luminosity is taken to be inversely proportional
to the binaries luminosity, implicitly assuming that all donors supply an
identical amount of mass.  Still, the method has the advantage of rapidly
exploring a wide parameter space, as well as providing analytic relations
between certain parameters that would not be immediately obvious from
numerical population synthesis.

A comparison of the results from \citet{wu:2001} is complicated by the
fact that no easy way exists to adjust his results to a
SFR. \citeauthor{wu:2001} presents a number of luminosity functions
representing the distribution of XRBs at a given epoch, but no
normalization is possible. Nevertheless, it is interesting to note
that the bulk of the cumulative luminosity functions shown by
\citet{wu:2001} display a distinct turnover at a luminosity of around
$6\times10^{37}~\ergs$, a turnover which is also clearly visible in
many of the luminosity functions resulting from our own simulations,
and roughly one-third of the Eddington luminosity of a typical neutron
star. As more work making use of this technique appears, a better
comparison of these two dramatically different approaches to
population synthesis may be possible.

\subsection{Semi-Analytical Calculation by Ghosh \& White}

\citet{ghosh:2001} make use of a related methodology through which
they use X-ray survey data to infer the long-term evolution of cosmic
SFRs. They also establish a system of differential equations
describing the population of X-ray binaries, though their approach is
more rooted in experimental data, and is couched in terms of a real
SFR, as opposed to the parameterized version found in \citet{wu:2001}.

The goal of this study is quite different from our own, and is focused
on the X-ray luminosity evolution over time scales of a Hubble
time. Our work is in rough agreement in a qualitative sense; that is,
the relative contribution of the various XRB species is similar, and
the shape of the luminosity curves are analogous. We differ in one
major prediction, however.  \citeauthor{ghosh:2001} predict that the
LMXB population will not become a significant contributor to the total
X-ray luminosity until several Gyr have passed; in other words, when
the companion stars either begin evolving off the main sequence, or
lose sufficient orbital energy to magnetic braking and/or
gravitational radiation. Our prediction is that a large number of LMXB
systems will become active in the first 200 Myr or so after a vigorous
star formation episode; e.~g., panels (e) and (f) on any of
Figures~\ref{fig:sal_lowq_bh}--\ref{fig:ms_flatq_ns}. These systems
are survivors of a stage of common-envelope evolution, which shrank
the orbital radius and permitted Roche-lobe overflow to occur several
billion years before it would otherwise be
possible. \citeauthor{ghosh:2001} do not consider the effect of
common-envelope evolution in their equation set, and so do not predict
this feature.

\section{Comparison With Observations\label{model:observations}}

We close by comparing some of our major predictions with
observations. The computed quantities that are readily comparable to
observations are the integrated X-ray luminosity of the population and
the general shape of the luminosity function and its evolution.
Further and more detailed tests must take the form of case studies,
where the X-ray properties of well-studied star-forming systems are
compared with customized simulations that make use of the specific
properties of those systems (such as the SFR, the age of the
starburst, and the star-formation history).

Recently, \citet{ranalli:2003} extended the well-known correlation
between a galaxy's far-infrared (and radio) luminosities and the
underlying SFR to the 2--10~keV energy regime \citep[an earlier form
of this correlation was also presented by][]{helfand:2001}. Making use
of and {\it ASCA} {\it BeppoSAX} observations of nearby actively-star
forming galaxies, they proposed that the SFR could be related to the
2--10~keV luminosity of the host galaxy by
\begin{equation}
{\rm{SFR}} =  
2.0\; \left( {L_{\rm 2-10\;keV}\over 10^{40}~\ergs} \right)~\Msunyr\; .
\end{equation}
For our simulated galaxy experiencing an SFR of 10 \Msunyr, we found
that the peak hard X-ray luminosity, reached at the time star
formation ceases and sustained for several tens of Myr, was around
$4\times10^{40}$ \ergs. As discussed earlier, this result is largely
independent of the choice of IMF or binarity fraction. By the above
relation, such a galaxy should have an underlying SFR of 8 \Msunyr,
which is in good agreement with the actual rate used in the
simulation. This reinforces our original claim that the principal
output of our simulations is robust.

\citet{grimm:2003} show a number of cumulative luminosity functions
based upon recent \emph{Chandra} and ASCA observations of nearby
starbursts. The cumulative luminosity functions of twelve noted
starbursts were fitted to a relation describing the fall off of the
luminosity function at the high end. They derive a best-fit power-law
index of $\beta=0.6$, which can be compared to our results in
Table~\ref{tbl:lfs}. \citet{grimm:2003} do not discuss the variation
in this index with time (as the stellar population evolves), thus we
compare their results to our cumulative luminosity function slopes at
early times as most of the objects in their study are still actively
star-forming. While still slightly steeper than those of
\citet{grimm:2003}, our index at 10 Myr, especially for the
Miller-Scalo IMF, is in agreement with the observations. It should
also be noted that the power-law indices measured for the individual
galaxies have a substantial spread to them and are in general
agreement with our predicted range of values. Similar measurements for
smaller sets of galaxies \citep{eracleous:2002, kilgard:2002}, are in
closer agreement with our steeper cumulative luminosity function. Our
cumulative luminosity functions show the correct behavior, as they
become both steeper and fainter with increasing time. This point is at
the heart of the observed difference between the luminosity functions
of spiral galaxies and ellipticals \citep{zezas:2002}. As well,
elliptical galaxies are generally more massive than the systems
generated by our simulations, so that the chance of finding rare,
exceptionally luminous sources (i.~e., with black hole primaries
descended from the very high end of the IMF) increases
dramatically. Such sources would affect the slope of the luminosity
function, but they are unlikely to be predicted by our simulations.

\acknowledgements 

This work was supported by NASA through grant GO0-1152A,B from the
Smithsonian Astrophysical Observatory. MSS acknowledges additional
support from the Zaccheus Daniel Foundation.  SS acknowledges support
from NSF grant PHY-0203046 and from the Center for Gravitational Wave
Physics, which is supported by the NSF under cooperative agreement
PHY~01-14375.  We thank Vicky Kalogera Krzys Belczynski for very
useful discussions.

\clearpage


\begin{deluxetable}{lcrrr}
\tablewidth{0in}
\tablecolumns{5}
\tablecaption{Normalization Parameters for Starburst Simulations\label{table:params}}
\tablehead{ 
\colhead{Simulation} & 
\colhead{$\bar{m}$} & 
\colhead{} & 
\colhead{} & 
\colhead{$n$} \\
\colhead{Parameters} & 
\colhead{$( = \bar{M}/\Msun)$} & 
\colhead{$\bar{q}$} & 
\colhead{$w$} & 
\colhead{(Myr$^{-1}$)}
}
\startdata
Sal/flat $q$ & 0.39 & 0.500 & 145.5 & 71300\\
Sal/low $q$  & 0.39 & 0.386 & 145.5 & 74700\\
M-S/flat $q$ & 0.60 & 0.500 &  69.2 & 96300\\
M-S/low $q$  & 0.60 & 0.386 &  69.2 & 101000\\
\enddata
\tablecomments{ Columns: (1)
simulation identifier, which shows both the mass function and mass ratio
distribution used, (2) mean mass of a star with
the given IMF, (3) mean mass ratio of a companion to the
primary, (4) conversion factor giving the total number of stars formed for
every star formed above 4~\Msun, (5) the number of binary systems
generated in each 1 Myr interval to give an SFR of $10~\Msunyr$.}
\end{deluxetable}

\begin{deluxetable}{lrrrrr}
\tablecolumns{6}
\tablewidth{0in}
\tablecaption{Power-Law Index of Model Cumulative Luminosity Function at Five Epochs\label{tbl:lfs}}
\tablehead{ 
\colhead{Model} &
\colhead{10 Myr} &
\colhead{20 Myr} &
\colhead{50 Myr} &
\colhead{100 Myr} & 
\colhead{200 Myr} 
}
\startdata
M-S / flat $q$ & 0.8--1.3 & 1.1--1.7 & 0.7--1.3 & 3.3--4.4 & 2.0--3.6 \\
M-S / low $q$  & 0.8--1.1 & 1.1--1.5 & 1.1--1.8 & 2.9--5.7 & 2.9--5.7 \\
Sal / flat $q$ & 1.2--2.5 & 1.3--1.9 & 1.4--3.6 & 2.9--5.7 & 2.9--5.0 \\
Sal / low $q$  & 0.9--1.5 & 1.1--1.5 & 0.9--1.6 & 2.9--4.4 & 2.2--3.6 \\
\enddata
\tablecomments{The measured ranges of the slope of a series of
model luminosity functions. The model column refers to the assumed IMF and
whether the mass ratio distribution is flat, or skewed towards low-mass
companions.}
\end{deluxetable}

\clearpage

\end{document}